\documentclass[aip,
jmp,%
 amsmath,amssymb,
preprint,%
]{revtex4-1}
\usepackage{graphicx}
\usepackage{dcolumn}
\usepackage{color}
\begin{document}

\newtheorem{lemma}{Lemma}[section]
\newtheorem{pro}[lemma]{Proposition}
\newtheorem{cor}[lemma]{Corollary}
\newtheorem{theorem}[lemma]{Theorem}

\bibliographystyle{unsrt}

\newcommand{\cierto}[1]{\fbox{#1}\label{#1}}

\newcommand\bu{\boldsymbol u}\newcommand\bU{\boldsymbol U}
\newcommand\ba{\boldsymbol a}\newcommand\bb{\boldsymbol b}
\newcommand\bn{\boldsymbol n}\newcommand\bk{\boldsymbol k}
\newcommand\bx{\boldsymbol x}\newcommand\by{\boldsymbol y}
\newcommand\bz{\boldsymbol z}
\newcommand\bv{\boldsymbol v}\newcommand\bV{\boldsymbol V}
\newcommand\bw{\boldsymbol w}\newcommand\bW{\boldsymbol W}
\newcommand\be{\boldsymbol e}\newcommand\br{\boldsymbol r}
\newcommand\bo{\boldsymbol 0}\newcommand\boo{\boldsymbol o}
\newcommand\bOO{\boldsymbol O}

\newcommand\btau{\boldsymbol \tau}
\newcommand\bh{\boldsymbol h}\newcommand\bg{\boldsymbol g}

\newcommand\bcalT{\boldsymbol{\cal T}}
\newcommand\bcalL{\boldsymbol{\cal L}}
\newcommand\bPhi{\boldsymbol{\Phi}}
\newcommand\bvarphi{\boldsymbol{\varphi}}
\newcommand\bOmega{\boldsymbol{\Omega}}

\newcommand\ip{\boldsymbol\cdot}\newcommand\dip{\boldsymbol :}
\providecommand\bnabla{\boldsymbol{\nabla}}

\newcommand\er{\mbox{\rm e}}\newcommand\ir{\mbox{\rm i}}

\newcommand\red{\color{red}}
\newcommand\blue{\color{blue}}
\newcommand\green{\color{green}}
\newcommand{\beqn}{\begin{equation}}
\newcommand{\eeqn}{\end{equation}}
\newcommand{\beqnarr}{\begin{eqnarray}}
\newcommand{\eeqnarr}{\end{eqnarray}}

\normalsize
\baselineskip=16truept
%\numberwithin{equation}{section}

\title[Optimal streaks]{Optimal  streaks in a Falkner-Skan boundary layer}
\author{Jos\'e J. S\'{a}nchez-\'{A}lvarez}
\author{Mar\'\i a Higuera}
\author{Jos\'e M. Vega\footnote{Corresponding author}}
\affiliation{E. T. S. I. Aeron\'auticos, Universidad Polit\'ecnica de Madrid,\\
Plaza Cardenal Cisneros 3, 28040 Madrid, SPAIN}

\begin{abstract}
 This paper deals with the
optimal streaky perturbations (which maximize the perturbed energy growth) in a
wedge flow boundary layer.  These three dimensional perturbations are
governed by a system of  linearized boundary layer equations
 around the  Falkner-Skan base flow.
Based on an asymptotic analysis of this system near the free stream and the
leading edge singularity, we show that  for acute wedge semi-angle,
all solutions converge  after a streamwise transient
to a single streamwise-growing solution of the linearized
equations, whose initial condition near the leading edge is given by
an eigenvalue problem first formulated in this context by Tumin (2001).
Such a solution may be regarded as  a streamwise
evolving most unstable streaky  mode,
in analogy with the usual eigenmodes in strictly parallel flows,
and shows an approximate self-similarity, which was partially known
and is completed in this paper.
An important consequence of this result is that the optimization
procedure based on the adjoint equations
heretofore used to define optimal streaks is not necessary.
Instead, a simple low-dimensional optimization process
is proposed and used to obtain optimal  streaks.  Comparison with
previous results  by Tumin and Ashpis (2003) shows an  excellent agreement.
 The unstable streaky mode exhibits transient growth if the
 wedge semi-angle
is smaller than a critical value that is slightly larger than $\pi/6$,
and decays otherwise. Thus the cases of right and obtuse wedge semi-angles exhibit
less practical interest, but they show a qualitatively different behavior, which
is briefly described to complete the analysis.
\end{abstract}
\maketitle
\def\etal{\text{et al.}}
\def\tand{\text{\quad and\quad}}
\def\at{\text{\quad at\quad}}
\def\cc{\hbox{c.c.}}
\font\ochob=cmbx8
\font\ocho=cmr8
\def\tru#1{\cierto{#1}\raise2truemm\hbox{\sc{#1}}}
\def\rin{{\rm in}}
\def\at{{\rm at}}
\def\and{{\rm and}}
\def\for{{\rm for}}
\def\cc{{\rm c.c.}}

\def\HOT{{\rm HOT}}
\def\Rey{{\rm Re}}

\allowdisplaybreaks

%####################################################################
%####################################################################
\section{Introduction \label{sec:intro}}
%####################################################################
%####################################################################
Internal streaks (or Klebanoff modes \cite{Klebanoff71})
of a two-dimensional laminar boundary layer
denote  three-dimensional   disturbances
that evolve slowly streamwise and show a fast oscillation
(with a wavelength comparable to the boundary layer thickness);
see Refs.\cite{Westin94, Alfredsson96, Matsubara01} for some recent experiments.
The linear stability analysis \cite{Ellingsen75, Landahl80} demonstrated
that three dimensional  perturbations with streaky
structure are supported in inviscid shear flows,
whose energy grows   algebraically in time. This  initial
 growth together with a subsequent  decay due to
 viscous dissipation
is known as transient growth \cite{Schmidt07}.
Streaks can  be forced either internally,
 from obstacles near the leading edge or externally, from perturbations
 in the free stream, and generally interact
 with the transversal Tollmien-Schlichting modes,
 either enhancing \cite{KlebanoffTS62} or
 delaying\cite{CossuB02} transition to turbulence, depending on the streak amplitude.
  In the former case, the effect is known as bypass transition \cite{Morkovin84}.

Luchini~\cite{Luchini1996} proposed in the scope of  the Blasius boundary layer an
analytical description of streaky  perturbations  with a
small (compared to $1/\delta$, where $\delta$ is the boundary layer thickness)
spanwise wavenumber, which is equivalent to the
limit of small distance to the leading edge.
 The extension of this analysis to the Falkner-Skan \cite{Schlichting68}
profile was made by
Tumin~\cite{Tumin2001}, who   derived an
eigenvalue problem describing the velocity components
of the streaky perturbation. He computed the
largest eigenvalue of this problem in  terms of the
wedge angle and  found that the
unbounded growth is suppressed as the angle exceeds a
threshold value, which is slightly larger than $\pi/3$.

An important issue related  to streaky
perturbations is to find the initial disturbance that
maximizes the  perturbed energy streamwise growth;
the resulting streamwise evolving streak is known as the optimal streak.
For strictly parallel flows, an explicit computation of
disturbances resulting in the maximum transient growth
is possible by optimizing over the eigenmodes of the Orr-Sommerfeld
operator, as has been done
for a number of particular parallel flows,  both in the contexts of
temporal\cite{Hultgren81,Boberg88,Gustavsson91,Butler92,Henningson93,Reddy93,Trefethen93}
and spatial stability\cite{TuminR01}.

For the non-parallel case, in the scope of
Blasius boundary layer, Luchini~\cite{Luchini2000}
and Andersson et al.~\cite{AnderssonBH99}
calculated optimal streaks using a method
based on the adjoint formulation, which is commonly
employed in
optimal-control problems for distributed parameter systems.
Using this method, Tumin and  Ashpis \cite{TuminAshpis2003}
computed the optimal streaks  in
a Falkner-Skan boundary layer and showed the effects of the spanwise wave number
and the wedge angle on transient growth.
In particular, they found that an adverse pressure gradient
increases the amplification whereas a favorable
presure gradient has the opposite effect (in accordance with the previous asymptotic result by
Tumin~\cite{Tumin2001}). Levin and Henningson \cite{Levin2003} also studied the optimal
disturbance in a Falkner-Skan base  flow,
obtaining  results that were consistent with those
 by Tumin and Ashpis~\cite{TuminAshpis2003}.

In a recent paper\cite{HigueraV09}, two of the authors  analyzed streaks
in a Blasius boundary layer and showed that, after an initial
transient, they approach a unique (up to a constant factor) `mode',
which was called the unstable streaky mode. This mode
is calculated from the streamwise
  evolving linearized equations
with well-defined initial conditions near the leading edge, which  are given by
the first eigenmode of the eigenvalue problem
first formulated by Luchini~\cite{Luchini1996}.
  The unstable streaky mode provides the optimal streaks
when the initial conditions are taken
 sufficiently close to the leading edge (say, $x\leq10^{-3}$),
making it unnecessary the optimization process in this limit. In fact, the asymptotic
value of the optimal spanwise wavenumber was found to be 0.484, which was slightly
different to  its counterpart (0.45) calculated by Luchini and Andersson et al. The latter
 corresponds to an only moderately small value of the initial streamwise coordinate,
$x\sim0.01$. The analysis in
Ref.\cite{HigueraV09} was based on three main
 ingredients:
\begin{enumerate}
 \item  The exact self-similarity of the boundary
 layer, which allows to eliminate the wavenumber from the formulation.
 \item The modal structure  near the
 leading edge, which allows to expand any initial condition into a complete system
       of  eigenmodes.
 \item An asymptotic analysis of the behavior of the solutions  near
 the free stream. This was essential to both understanding the structure of the
 solutions and formulating a quite efficient numerical scheme, which provided streaks
     with initial conditions at a section extremely close to the leading edge.
\end{enumerate}
In addition, the approximate self-similarity already detected
 experimentally in Ref.\cite{Westin94} and confirmed in Ref.\cite{Luchini1996}
 (namely, the wall-normal profile of the streamwise velocity component of
 optimal streaks, rescaled with its maximum, remains constant streamwise)
 was completed noting that the wall-normal profile of a  certain
 combination of the cross flow velocity components shows the same
 property as the streamwise velocity component.  A part of the present
 paper is an extension of this previous work to the Falkner-Skan boundary layer.
 In addition, advantage will be taken of the low dimensional nature of streaks
  to develop a quite efficient and simple method to calculate optimal streaks.
  This method consists in optimizing the perturbed energy gain in a low-dimensional
  solution manifold spanned by a few  solutions
  of the streamwise evolving problem.

The  remaining of the paper starts with the  formulation of the problem,
in section \ref{sec:Formulation}, where the asymptotic behaviors near the free
stream and the leading edge are analyzed, in subsections  \ref{sec:ZetaInfty}
and  \ref{sec:leading-edge}, respectively. The streamwise evolution of streaks
is considered in
section \ref{sec:streamevolution}, where a quite efficient numerical scheme
is presented that relies on the
previous asymptotic results; the modal nature of streaks is considered in subsection
\ref{sec:usm}. Optimal streaks are studied  in
section \ref{sec:optimalstreaks}, where the above mentioned optimization
method is presented, in subsection
\ref{sec:method}. The body of the paper deals with acute wedge semi-angles;
the cases of right and obtuse  semi-angles are considered
in an appendix, at the end of the paper.
 The paper ends with some concluding
remarks, in section \ref{sec:Concluding}.

%####################################################################
%####################################################################
\section{General Formulation \label{sec:Formulation}}
\begin{figure}
\includegraphics[width=10cm]{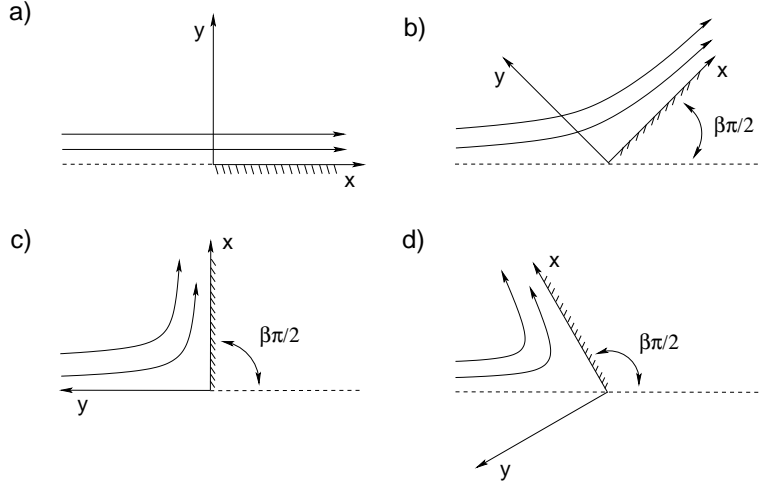}
 %\center{\includegraphics[width=10cm]{sketchBL.eps}}
 \caption{ The symmetric flow past
a wedge  with angle $\pi\beta$, in the cases (a) $\beta=0$ (Blasius flow),
(b) $0<\beta<1$, (c) $\beta=1$ (stagnation flow),
and (d) $1<\beta<2$. \label{fig1}}
\end{figure}

%####################################################################
%####################################################################
We  consider the  high-Reynolds-number flow incidenting in a wedge
with  semi-angle  $\beta\pi/2$ (Fig.\ref{fig1}),
 where the  parameter $\beta$ is named after
Hartree\cite{Hartree37}. In particular, we study the linear
stability  of the resulting  boundary layer flow, under  streaky
 perturbations  that
exhibit a  spanwise  period comparable to the thickness of the boundary layer,
$\delta^*=L^*/\sqrt{\text{Re}}$,
 where $L^*$  is the wall wise length of the portion of the wedge
 under consideration  and $\text{Re}=u^*L^*/\nu\gg1$ is the Reynolds
number based     on  $L^*$ and the free stream velocity $u^*$.
Nondimensionalization is made according to the usual
boundary layer  approximation,
using the following units: $L^*$ and $u^*$ for the streamise spatial
coordinate $x$ and velocity $u$, respectively, and
 $\delta^*$ and $u^*/\sqrt{\text{Re}}$ for the wall-normal and spanwise
coordinates $(y,z)$ and velocities $(v,w)$, respectively;
the pressure is scaled with $\rho^*(u^*)^2/\sqrt{\text{Re}}$.
%%%%%%%%%%%%%%%%%%%%%%%%%%%%%%%%%%%%%%%%%%%%%%%%%%%%%%%%
%                    FIGURA
%%%%%%%%%%%%%%%%%%%%%%%%%%%%%%%%%%%%%%%%%%%%%%%%%%%%%%%%
%%%%%%%%%%%%%%%%%%%%%%%%%%%%%%%%%%%%%%%%%%%%%%%%%%%%%%%%
%%%%%%%%%%%%%%%%%%%%%%%%%%%%%%%%%%%%%%%%%%%%%%%%%%%%%%%%
 The  two-dimensional, symmetric base flow,
$(u, v, w,p)=(u_b, v_b, 0,p_b)$, is  the
Falkner-Skan\cite{FalknerS31}
similarity solution
 \begin{alignat}{1}
& u_b(x,\zeta)=x^{\frac{\beta}{2-\beta}} F'(\zeta),
\quad v_b(x,\zeta)=-\frac{F(\zeta) +
(\beta-1) \zeta F'(\zeta)}{(2-\beta)g(x)},\label{bflow}
\end{alignat}
 where $\zeta$ is the selfsimilar, wall normal coordinate
\begin{equation}
\zeta=y/g(x),\quad\text{with }\quad  g(x)=x^{\frac{1-\beta}{2-\beta}}, \label{blv}
\end{equation}
  and
the streamfunction $F$ is the solution of the Falkner-Skan equation
\begin{alignat}{1}
& F'''+[FF''+\beta (1-F'^2)]/(2-\beta)=0\quad\text{in}
\quad 0<\zeta<\infty,\label{FSec1}\\
&F(0)=F'(0)=0,\quad F'(\infty)=1.\label{FSec2}
\end{alignat}
 The Falkner-Skan solution  includes the cases (Fig.\ref{fig1}) of
 flat plate boundary layer $\beta=0$, flow past an acute
wedge ($0<\beta< 1$),  stagnation flow towards a flat plate ($\beta=1$), and
 flow in an obtuse wedge ($1<\beta< 2$); negative values of $\beta$
involve unphysical reverse flow and are only
of academic  interest.  Note that the scaled boundary layer
thickness is proportional to $g(x)$, which invoking (\ref{blv}) means that it
grows streamwise if $\beta<1$ and decays if $\beta>1$, while at $\beta=1$
the boundary layer flow is parallel. The remaining of the paper will concentrate
in the case of acute wedge angle. The case $\beta\geq1$ does not promote
transient growth and will be briefly analyzed in the Appendix, at the end
of the paper.

Perturbations are considered linearizing  around the
 base flow
 and decomposing
   in normal modes as
   \beqn
   (u-u_b,v-v_b,w, p-p_b)= \left(U(x,\zeta), \frac{g(x)V(x,\zeta)}{x},
   \frac{\ir g(x)W(x,\zeta)}{x},
   \frac{P(x,\zeta)}{x\sqrt{\text{Re}}}\right)\er^{\ir \alpha z},\label{expansion}
   \eeqn
  to obtain the following {\it linearized boundary layer} (LBL) equations
\begin{alignat}{1}
&{x}\partial_{{x}}{U}=
 \frac{(1-\beta)\zeta }{2-\beta}\partial_{\zeta} {U}
 - \partial_{\zeta}{V}-
\alpha g( x) W,\label{mlbl10} \\
&{x}F'\partial_{{x}}{U}=
\partial_{\zeta \zeta}U +  \frac{F}{2-\beta}\partial_\zeta{U}-
\frac{\beta F'+(\beta-1)\zeta F''+(2-\beta)\alpha^2g({x})^2}{2-\beta}{U}
-F''{V},\label{mlbl20}\\
& {x}F'\partial_{\hat{x}}{V}=
\partial_{\zeta \zeta} V+\frac{F}{2-\beta}\partial_\zeta{V}
+ \frac{(\beta-1)[F+(2\beta-1)\zeta F'+(\beta-1)\zeta^2F'']}{(2-\beta)^2}{U}\nonumber\\
&\hskip4cm+
\frac{(\beta+1)F'+(\beta-1)\zeta F''-(2-\beta)\alpha^2g({x})^2}{2-\beta}{V}
-\partial_\zeta{P},\label{mlbl30}\\
&{x}F'\partial_{{x}}{W}=
\partial_{\zeta \zeta} W+\frac{F}{2-\beta}\partial_\zeta{W}+
\frac{F'-(2-\beta)\alpha^2g({x})^2}{2-\beta}{W}-\alpha g({x}) P,\label{mlbl40}\\
&U=V= W =0 \quad\text {at}\quad \zeta=0\text{ and }
\infty,\quad  P=0\quad\text {at}\quad  \zeta=\infty,\label{mlblcc0}
\end{alignat}
where $\zeta$ is the selfsimilar wall normal coordinate defined in eq.(\ref{blv}) and
$\partial_x,\partial_y,\cdots$ denote hereafter partial derivatives. Note that
the boundary conditions are no-slip at the plate and vanishing at the
infinity in the wall-normal  direction.

If $\beta=1$, then $g(x)=1$ and the right hand sides of
eqs.(\ref{mlbl10}-\ref{mlbl40}) are independent of $x$. Otherwise,
the spanwise wavenumber can be eliminated from the formulation
using the scaling
\begin{alignat}{4}
& (\hat x,\hat y,\hat{u}_b,\hat{v}_b) =(\alpha^{\frac{2-\beta}{1-\beta}},\alpha y,\alpha^{\frac{\beta}{1-\beta}} u_b,v_b/\alpha),
 \quad (\hat U,\hat V,\hat W,\hat P)=\alpha^{\frac{1}{1-\beta}} (U,V,W, P).\label{b19}
\end{alignat}
In addition, as in Ref.\cite{HigueraV09}, we use  the new cross flow variable
\beqn
\hat H=\hat V+\hat W,\label{b50}
\eeqn
which anticipating results in
subsection \ref{sec:ZetaInfty} below,  converges quite fast to zero as $\zeta\to\infty$.
Thus, we substitute the spanwise
velocity component by this new variable in the formulation. Using all these,
eqs.(\ref{mlbl10})-(\ref{mlblcc0}) are rewritten as
\begin{alignat}{1}
&\hat x\partial_{\hat x}\hat U=\frac{1-\beta}{2-\beta}\zeta\partial_\zeta \hat U
 -\partial_\zeta\hat V-g(\hat x)\hat V
 +g(\hat x) \hat H,\label{mlbl1} \\
&F'\hat{x}\partial_{\hat{x}}\hat{U}=
\partial_{\zeta \zeta}\hat U +  \frac{F}{2-\beta}\partial_\zeta\hat{U}-
\frac{\beta F'+(\beta-1)\zeta F''+(2-\beta)g(\hat{x})^2}{2-\beta}\hat{U}
-F''\hat{V},\label{mlbl2}\\
& F'\hat{x}\partial_{\hat{x}}\hat{V}=
\partial_{\zeta \zeta}\hat V+\frac{F}{2-\beta}\partial_\zeta\hat{V}
+ \frac{(\beta-1)[F+(2\beta-1)\zeta F'
+(\beta-1)\zeta^2F'']}{(2-\beta)^2}\hat{U}\nonumber\\
&\hskip4cm+
\frac{(\beta+1)F'+(\beta-1)\zeta F''
-(2-\beta)g(\hat{x})^2}{2-\beta}\hat{V}-\partial_\zeta\hat{P},\label{mlbl3}\\
&\hat xF'\partial_{\hat x}\hat H
=\partial_{\zeta\zeta}\hat H+\frac{F}{2-\beta}\partial_\zeta \hat H
+\frac{(\beta-1)[F+(2\beta-1)\zeta F'+(\beta-1)\zeta^2F'']}{(2-\beta)^2}\hat U\nonumber\\
&\hskip2cm+\frac{\beta F'+(\beta-1)\zeta F''}{2-\beta}\hat V
+\frac{F'-(2-\beta)g(\hat x)^2}{2-\beta}\hat H
-\partial_\zeta\hat P-g(\hat x)\hat P,\label{mlbl4}\\
&\hat U=\hat V=\hat H =0 \quad\text {at}\quad \zeta=0\text{ and }
\infty,\quad \hat P=0\quad\text {at}\quad  \zeta=\infty.\label{mlblcc}
\end{alignat}
 These equations will  referred to below as the {\it modified linear boundary layer}
(MLBL) equations.

%####################################################################
%####################################################################
\subsection{Asymptotic behavior near the  free stream ($\zeta >> 1$) for $\beta<1$\label{sec:ZetaInfty}}
%####################################################################
%####################################################################
As  $\zeta\rightarrow\infty $, the streamfunction of the Falkner-Skan base flow behaves as
\begin{equation}
F(\zeta) \sim \zeta- a_\beta + O(e^{-(\zeta -
a_\beta)^2/2}),\label{bfas}
\end{equation}
where the constant  $a_\beta$ depends on Hartree parameter $\beta$. Thus
 both $F'-1$ and $F''$  decay to zero
 exponentially fast and  the LBL equations
  (\ref{mlbl10})-(\ref{mlbl40}) can be greatly simplified.
  In particular, the streamwise momentum
  equation (\ref{mlbl20}) becomes
 \beqn
  {x}\partial_{{x}}{U}=\partial_{\zeta \zeta} U +
  \frac{\zeta- a_\beta}{2-\beta}\partial_\zeta{U}-
\frac{\beta+(2-\beta)\alpha^2 g({x})^2}{2-\beta}{U}.\label{mlblas1a}
  \eeqn
  This equation is unforced, which means that $U=0$.
  Using this, the remaining  LBL equations
  (\ref{mlbl10}), (\ref{mlbl30})-(\ref{mlbl40}) simplify to
\begin{alignat}{1}
&\partial_\zeta {V} =\alpha g({x}){W},\label{mlblas1}\\
&{x}\partial_{{x}}{V}=\partial_{\zeta \zeta}{V}
+\frac{\zeta- a_\beta}{2-\beta}\partial_\zeta{V}
+\frac{\beta+1-(2-\beta)\alpha^2g({x})^2}{2-\beta}{V}-\partial_\zeta{P},
\label{mlblas2}\\
&{x}\partial_{{x}}{W}=\partial_{\zeta \zeta} W
+\frac{\zeta- a_\beta}{2-\beta}\partial_\zeta{W}+
\frac{1-(2-\beta)\alpha^2g({x})^2}{2-\beta}{W}-\alpha g({x})P.\label{mlblas3}
\end{alignat}

 These equations can be solved in closed form as follows.
We first eliminate the pressure manipulating
(\ref{mlblas2}) and (\ref{mlblas3}) as usually,
to obtain the following equation for the streamwise vorticity
${\Omega}=\partial_\zeta W-\alpha g( {x}) {V}$
\begin{equation}
{x}\partial_{x}{\Omega}=\partial_{\zeta \zeta}{\Omega}
+\frac{\zeta-a_\beta}{2-\beta}\partial_\zeta{\Omega}
+\frac{2-(2-\beta)\alpha^2 g({x})^2}{2-\beta}{\Omega}.\label{vortix}
\end{equation}

This equation is also unforced, which means that $\Omega=0$, namely
\begin{equation}
\partial_\zeta W=\alpha g({x}) {V}.\label{mlblas5}
\end{equation}
Excluding divergent behaviors
 as $\zeta\to\infty$, eqs.(\ref{mlblas1}) and (\ref{mlblas5}) yield
 \begin{equation}
V=-W= V_\infty \er^{-\alpha g(x)(\zeta-a_\beta)},
\label{UVWPinfty}
\end{equation}
where  $V_\infty$ is a function of $x$  that
remains undetermined but will not be
necessary below. The pressure $P$ is readily obtained (and seen to behave as
$\zeta \er^{-\alpha g(x)(\zeta-a_\beta)}$) substituting
(\ref{UVWPinfty})  into (\ref{mlblas3}).

Since $\beta<1$,
$g(x)=x^{\frac{1-\beta}{2-\beta}}$ is small at small $\hat x$ and
the convergence of
$V$, $W$, and $P$ to the
final free stream state $U=V=W=P=0$ is quite slow (see Fig.\ref{fig3} below),
 which explains
the difficulties encountered in former
numerical treatments\cite{AnderssonBH99,Luchini2000,TuminAshpis2003}
of (\ref{mlbl10})-(\ref{mlblcc0}) that
did not take into account this behavior.
The streamwise velocity component $U$ instead, converges quite fast
to zero (see Fig.\ref{fig3} below),  as $\er^{-(\zeta- a_\beta)^2/2}$.
Equation (\ref{UVWPinfty}) shows that the same happens with the  quantity  (cf (\ref{b50}))
 \begin{equation}
 H=V+W,\label{vH}
 \end{equation}
 as anticipated above. Note
  using eqs.(\ref{mlblas1}), (\ref{mlblas5}), and (\ref{UVWPinfty})
 that $H$  coincides with the streamwise vorticity at large $\zeta$.

%##############################################################################
%##############################################################################
\subsection{Behavior near leading edge (${x}\ll 1$)
in the case $\beta<1$ \label{sec:leading-edge}}
%##############################################################################
%##############################################################################
Assuming that the solution behaves as a power of $\hat{x}$
 at small $\hat x$,
the relevant behavior is given by an
eigenvalue problem first formulated and solved by
Tumin \cite{Tumin2001}, whose formulation
 is re-interpreted here. This will be done
taking into account the asymptotic behavior as $\zeta\to\infty$
encountered in last sub-section, which
in conjunction with the Tumin scaling suggests the ansatz
\begin{equation}
 (\hat{U}, \hat{V},   \hat H, \hat {P}) \sim \hat x^{1-\lambda}
 (\tilde U,\tilde V,\tilde H/g(\hat x),\tilde P)
 e^{-g(\hat{x})(\zeta -  a_\beta)}.\label{UVHPx0}
\end{equation}
 Substituting this into the MLBL-equations and neglecting
 $O(g(\hat x))=O(\hat x^{\frac{1-\beta}{2-\beta}})$-terms
  yields the following eigenvalue problem
\begin{alignat}{3}
& (1-\lambda)\tilde  U + \frac{(\beta-1)\zeta}{2-\beta}\tilde U'
 +\tilde V' -\tilde H = 0, \label{mT1}\\
&\tilde U''+\frac{F}{2-\beta} \tilde U'
+\frac{(1-\beta)\zeta F''+[(2-\beta)\lambda-1]F'}{2-\beta}\tilde U
-F''\tilde V =0,\label{mT3} \\
& \tilde V''+\frac{F}{2-\beta} \tilde{V}'+
   \frac{(\beta-1)[F+(2\beta-1)\zeta F'
+(\beta-1)\zeta^2F'']}{(2-\beta)^2}\tilde{U}\nonumber\\
&\hskip3cm+\frac{[(2-\beta)\lambda+2\beta-1]F'+(\beta-1)\zeta F''}
{2-\beta}\tilde{V}=\tilde{P}',\label{mT3a}\\
&\tilde H''+\frac{F}{2-\beta}  \tilde H' +\lambda F' \tilde H =0,\label{mT4}\\
&\tilde U=\tilde V=\tilde H=0, \quad \text{at}\quad\zeta=0,
\quad \tilde U,\tilde V,\tilde H\rightarrow 0 \text{ as } \zeta\to
 \infty,\label{mT5}
\end{alignat}
 which coincides with that in Ref.\cite{Tumin2001} except for the
fact that we are using the variable
$\tilde H=\tilde U+\tilde W$ instead of $\tilde W$.
This makes sense since the relevant
eigenfuncions are such that
$\tilde H=O\left(\er^{-(\zeta- a_\beta)^2/2}\right)$
 as $\zeta\to\infty$ (Fig.\ref{fig2}, right), which is consistent with the
asymptotic behavior of the variable $H$
(see Fig.\ref{fig3} below) as explained above,
but not with that of $\hat W$, whose decay is much slower, as
 $\er^{-g(\hat x)(\zeta-a_\beta)}$ (see eq.(\ref{UVWPinfty})).

%%%%%%%%%%%%%%%%%%%%%%%%%%%%%%%%%%%%%%%%%%%%%%%%%%%%%%%%
%                    FIGURA
%%%%%%%%%%%%%%%%%%%%%%%%%%%%%%%%%%%%%%%%%%%%%%%%%%%%%%%%
 \begin{figure}
 \includegraphics[width=6cm]{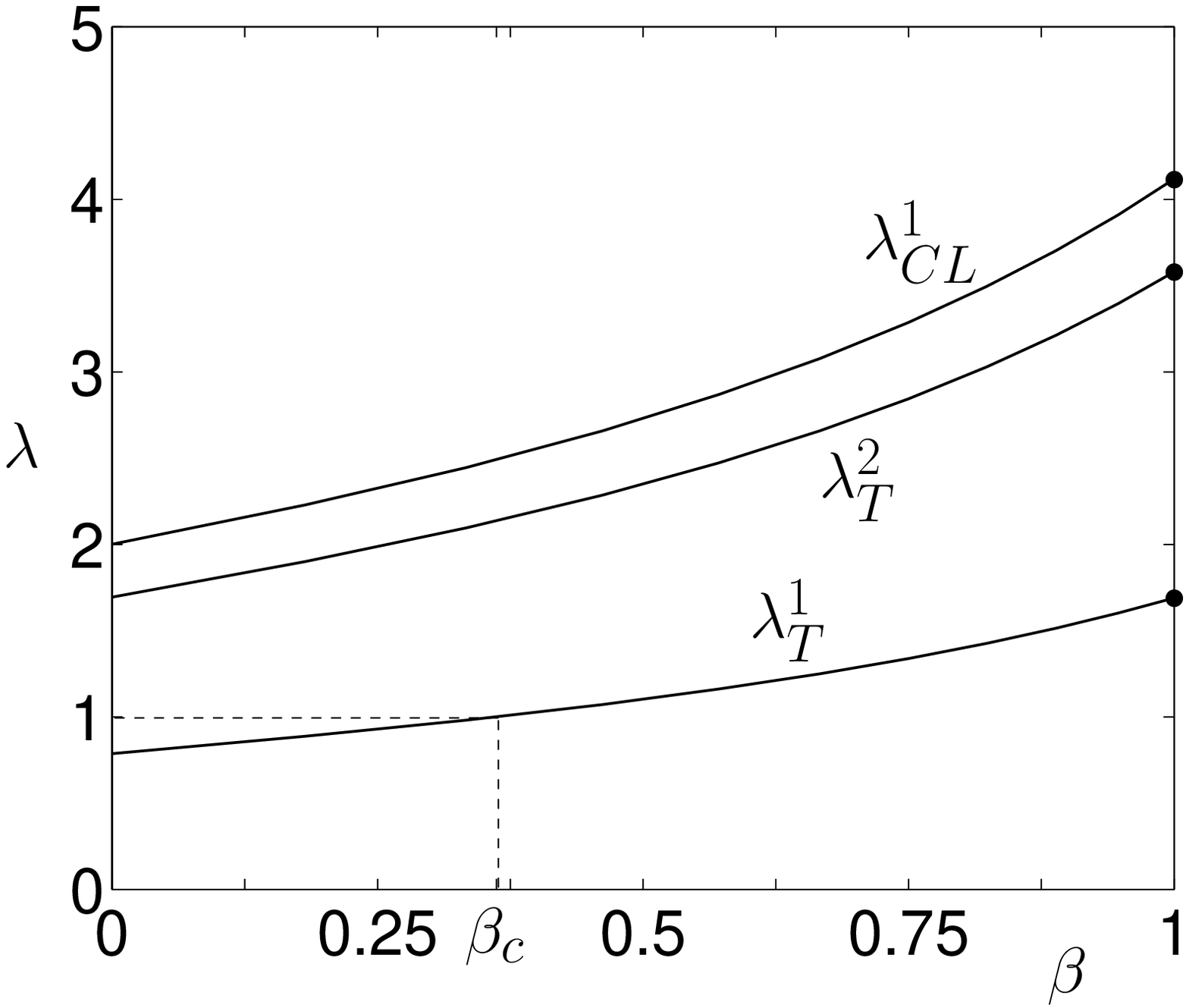}\includegraphics[width=6cm]{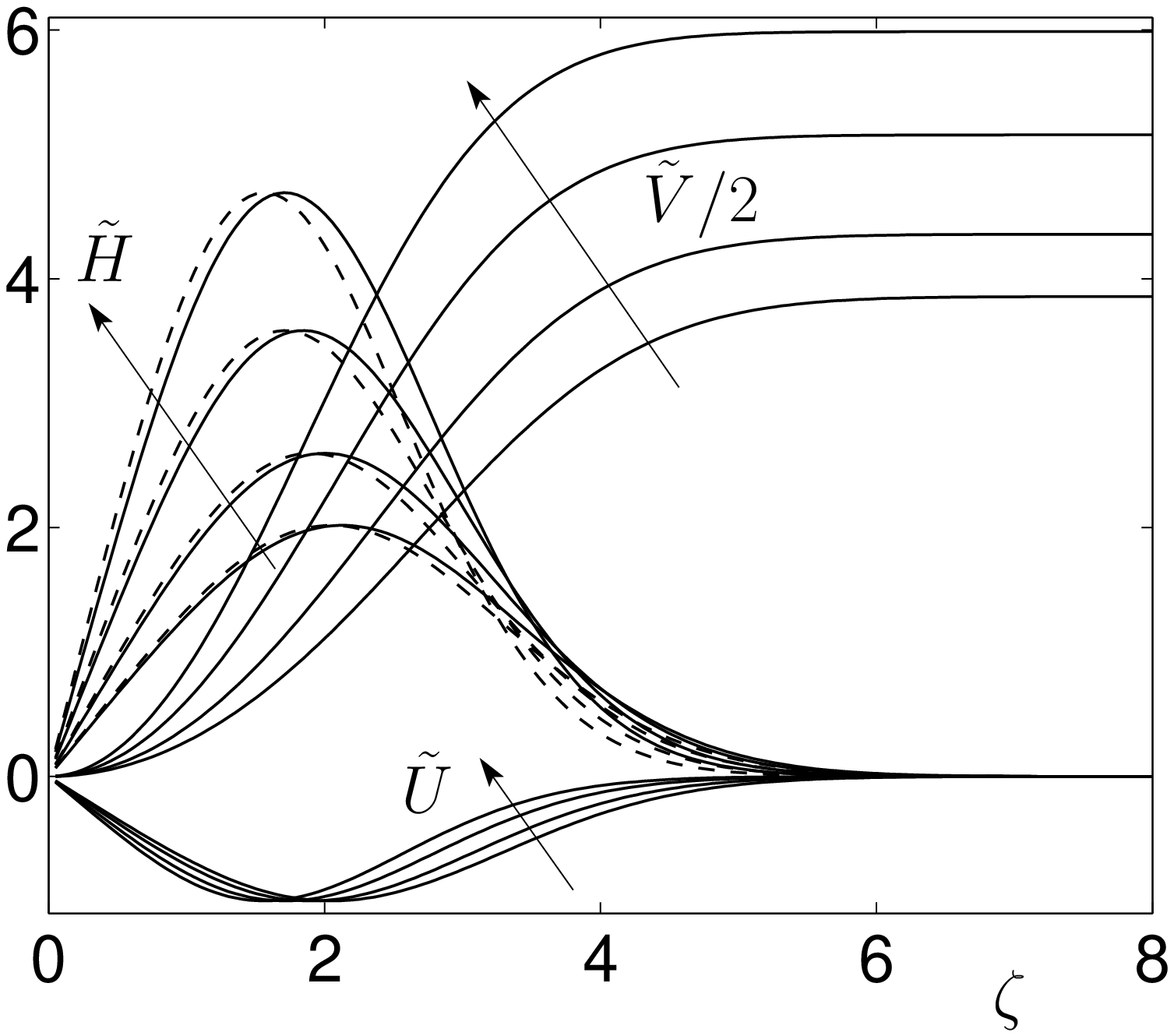}
   \caption{Left: The two smallest Tumin
 eigenvalues (subscript $T$) and the smallest
 Chen-Libby eigenvalue (subscript $CL$) in terms of  $\beta$,
  in the range $0\leq\beta\leq1$;
  the first Tumin eigenvalue is smaller than one
  if $0\leq\beta<\beta_c=0.362$. Plain circles for
  $\beta=1$ come from calculations in the Appendix.
  Right: $\tilde U$, $\tilde V$,
 $\tilde H$ (solid lines), and
 $\tilde U_0=-|\tilde H|_\text{max}\tilde U/|\tilde U|_\text{max}$
 (dashed lines) for the first Tumin mode and
  $\beta=0.1,0.2,0.353$ (that value considered
   in Ref.\cite{TuminAshpis2003}), and $0.5$; arrows
 indicate increasing $\beta$. \label{fig2}}
 \end{figure}
%%%%%%%%%%%%%%%%%%%%%%%%%%%%%%%%%%%%%%%%%%%%%%%%%%%%%%%%
%%%%%%%%%%%%%%%%%%%%%%%%%%%%%%%%%%%%%%%%%%%%%%%%%%%%%%%%

 Now,
eq.(\ref{mT4}) decouples from the remaining three
equations and yields the eigenvalue $\lambda$.
These eigenvalues will be called Tumin eigenvalues hereafter,
and are all positive. The two smallest eigenvalues
are plotted with solid line and indicated with
 the  subscript T in Fig.\ref{fig2}, left.
Note that the first eigenvalue is larger than 1
if $0\leq \beta<\beta_c= 0.362$,
which invoking eq.(\ref{UVHPx0}) means
that the associated streamwise and wall normal velocity
components exhibit algebraic growth; the spanwise
 component instead decays streamwise, as do all velocity
 components for the remaining eigenvalues. If
  $\beta>\beta_c$, all eigenvalues are
larger than one and all flow variables decay streamwise.
The eigenfunction components $\tilde U$,
$\tilde V$, and $\tilde H$ associated with the first Tumin
eigenvalue are plotted in Fig.\ref{fig2}, right
for various representative values of  $\beta$. Note
that   $\tilde U_0=-|\tilde H|_\text{max}\tilde U/|\tilde U|_\text{max}$
is always quite close to $\tilde H$, which means that $\tilde U$ and
$\tilde H$ are always almost linearly dependent.

 In order to obtain a complete system of
 eigenfunctions (to, e.g., set all possible initial conditions),
 a second eigenvalue problem must be considered, which
   results from the  scaling  (cf (\ref{UVHPx0}))
\begin{equation}
 (\hat{U}, \hat{V}, \hat H, \hat {P})
 \sim \hat x^{1-\lambda}(\tilde U,\tilde V,\tilde H,\tilde P)
  e^{-g(\hat{x})(\zeta -  a_\beta)}.\label{UVHPx1}
\end{equation}
Proceeding as above, a second eigenvalue problem results
that is a three-dimensional extension of
 its counterpart first considered by
Chen and Libby \cite{ChenLibby1968} in a two-dimensional setting.
The resulting  continuity and
spanwise momentum equations must be replaced by
\begin{alignat}{3}
&(1-\lambda)\tilde  U + \frac{(\beta-1)\zeta}{2-\beta}\tilde U'
+\tilde V' = 0, \label{ChL1}\\
&\tilde H''+\frac{F}{2-\beta}  \tilde H' +
   \frac{(\beta-1)[F+(2\beta-1)\zeta F'
+(\beta-1)\zeta^2F'']}{(2-\beta)^2}\tilde{U}\nonumber\\
&\hskip2cm+\frac{\beta F'+(\beta-1)\zeta F''}
{2-\beta}\tilde{V}+
\frac{(2-\beta)\lambda+(\beta-1)}{2-\beta}F' \tilde H=\tilde P',
\label{ChL4}
\end{alignat}
but the remaining equations are
still eqs.(\ref{mT3}) and (\ref{mT3a}),
as are the boundary conditions (\ref{mT5}).
 Thus, eqs.(\ref{mT3}) and (\ref{ChL1})
 are decoupled and provide the eigenvalues,
 called hereafter Chen-Libby eigenvalues.
These eigenvalues are all
larger than one (the smallest one is plotted
 in Fig.~\ref{fig2}, left) and thus they
 promote streamwise decay; they are also larger than
the first Tumin eigenvalue, which
provides the most dangerous behavior.

An important property of  the two eigenvalue problems considered above
 is that any arbitrary initial condition  of the MLBL equations,
$(\hat U_0,\hat V_0, \hat H_0,\hat P_0)$, can be
 expanded in terms of the associated modes.  In particular,
the component of any initial condition on the first
Tumin eigenmode will be relevant in section \ref{sec:streamevolution}. It
is given by
\begin{equation}
  a=\frac{\displaystyle\int_0^\infty
  e^{\frac{1}{2-\beta}\int_{0}^{\zeta}
 F d\eta} F'   \hat H_0 \tilde H_T d \zeta}
 {\displaystyle\int_0^\infty e^{ \frac{1}{2-\beta}
 \int_{0}^{\zeta}  F d\eta} F' \tilde H_T^2 d \zeta},
\label{amplmodeT}
\end{equation}
as obtained multiplying (with the $L_2$ inner product) the series expansion by
 the adjoint of the first Tumin mode,
 $(\tilde U_T^*,\tilde V_T^*,\tilde H_T^*,\tilde P_T^*)=(0,0, e^{\frac{1}{4-2\beta}\int_{0}^{\zeta}F d\eta}\tilde H_T,0)$, and
 taking into account that this is orthogonal to the remaining eigemodes. The expression above of the
 adjoint of the first Tumin eigenmode is obtained
  rewriting
eqs.~(\ref{mT1})-(\ref{mT4}), which provide the first Tumin mode, as
${\cal L}_1\tilde U_T+{\cal L}_2\tilde H_T=0$, ${\cal L}_3\tilde H_T=0$.
The adjoint of this problem is
\begin{equation}
 {\cal L}^*_1\tilde U_T^*=0, \quad {\cal L}^*_2\tilde U^*
 +{\cal L}_3^*\tilde H^*=0,\label{adjmodosTumin}
 \end{equation}
where ${\cal L}^*_j$ stands for the adjoint of the operator ${\cal L}_j$. Since $\lambda$ is not
an eigenvalue associated with the Chen-Libby problem,
the first equation of (\ref{adjmodosTumin})
 implies that $\tilde U_T^*=0$,  which means that $\tilde V^*_T=\tilde P^*_T=0$ and
 $\tilde H_T^*$ is given by ${\cal L}_3^*\tilde H_T^*=0$,
 where ${\cal L}_3^*$ is the adjoint of the operator
 defined by the left hand side of eq.(\ref{mT4}). Then,
 $\tilde H_T$ is readily found to be
 $\tilde H_T^*=e^{\frac{1}{4-2\beta}\int_{0}^{\zeta}F d\eta}\tilde H_T$
 if the usual  $L_2$
 inner product is used.

%################################################
%################################################
\section{Streamwise evolution of streaks for $\beta<1$\label{sec:streamevolution}}
%################################################
%################################################
The numerical  method to integrate the MLBL equations
(\ref{mlbl1})-(\ref{mlblcc}) is  an extension of
its counterpart developed  in Ref.\cite{HigueraV09}.
Since the basic steady state converges
 quite fast to its asymptotic value  as $\zeta \rightarrow \infty$,
the approximation  in subsection \ref{sec:ZetaInfty}
applies for moderately large values of
$\zeta$, say $\zeta>L_0=12$. In particular, eqs.(\ref{mlblas1})
and (\ref{mlblas5})  can be used, which excluding
divergent behaviors as $\zeta\to\infty$, lead to
\begin{equation}
\partial_\zeta\hat V+g(\hat x)\hat V=0.\label{mlblas5est}
\end{equation}
This equation is independent of all MLBL equations.
 If we substitute the wall normal momentum equation
 by this equation, the resulting system is readily seen
 to provide the  correct asymptotic behavior as
 $\zeta\to\infty$, analyzed in subsection \ref{sec:ZetaInfty}.
 Thus, we select the domain of integration $L=20>L_0=12$ and consider
 eqs.(\ref{mlbl1}), (\ref{mlbl2}), and (\ref{mlbl4})
  in $0<\zeta<L$, eq.(\ref{mlbl3})
 in $0<\zeta\leq L_0$, and eq.(\ref{mlblas5est})
 in $L_0<\zeta<L$. In addition, second order spatial
 derivatives in  the resulting system are
 discretized using second order centered differences; first
 order derivatives  of  $\hat V$ and $\hat P$
 are discretized using second order
 forward differences in
 eqs.(\ref{mlbl1}) and (\ref{mlblas5est})
 and second order backward differences in (\ref{mlbl3}); the latter
 are also used to discretize first order derivatives
 of $\hat U$ and $\hat H$ in all equations.
 This means (noting that both $\hat V$
 and $\hat P$ are discretized in $L_0<\zeta<L$ with
 forward differences)
 that no boundary conditions for $\hat V$ and
  $\hat P$ are needed at $\zeta=L$; instead,
 $\hat V(L)$ and $\hat P(L)$ are selected
 by the numerical code itself. Since both $\hat U$ and
 $\hat H$ decay extremely fast as $\zeta\to\infty$,
 the boundary conditions for  these at $\zeta=L$ are  $\hat U=\hat H=0$.
 The resulting equations
 can be written as $\hat x\textbf{\cal M}(\hat x)d\textbf{q}/dt
 =\textbf{\cal L}(\hat x)\textbf{q}$,
 where $\textbf{q}=(\hat U,\hat V,\hat H,\hat P)^\top$ is
 the joint flow vector and the matrices
 $\textbf{\cal M}$  and  $\textbf{\cal L}$
 result from the left and  right hand sides of the equations.
   The new logarithmic
 variable $s=log (\hat x/\hat x_0)$ is used
 to integrate the system from $\hat x=\hat x_0$ to
 $\hat x=1$,  discretizing $s$-derivatives  with
 second order forward differences, and using
 an implicit scheme to march in $s$.

The  resulting numerical scheme is very  efficient
because the numerical instabilities are excluded and the behavior
as  $\zeta\to \infty$  is well captured by (\ref{mlblas5est}).
The scheme is also robust
because results are  insensitive to both the choice of $L$ and $L_0$ (provided that $L>15$, $L_0>10$,
 and $L$ is somewhat large compared with $L_0$) and  the spatial and $s$-discretizations
 (provided that the latter be not too fine compared with the former since the mass matrix
 $\textbf{\cal M}$ is singular).
  Note that having solved the problem for $\zeta\rightarrow\infty$  has allowed to
describe accurately the flow very close of the leading edge ($\hat x\rightarrow 0$), without
 the need of using a extremely fine mesh  in $\zeta$,
which would have made the method  impractical at small $\hat x$.
A typical run is carried out in $10^{-9}<\hat x<1$ with $L_0=12$ and $L=20$,  considering
  $200$ equispaced $\zeta$-mesh points  and performing  400 $s$-steps, which  requires
20 CPU seconds using a Fortran90 code in a standard desktop computer.
%#################################################################
%#################################################################

\subsection{The most unstable streaky mode for $\beta<1$\label{sec:usm}}
 Let us define the most unstable streaky mode
 (MUSM) as that solution of the MLBL equations
that results from taking as initial conditions at $\hat x=\hat x_0\ll1$ precisely
that reconstructed via eq.(\ref{UVHPx0}) using the first Tumin eigenfunction defined by
 (\ref{mT1})-(\ref{mT5}), with $\lambda=\lambda_1^T$.
 Note that the MUSM is defined up to a constant factor, common to all flow
variables.
%#################################################################
%#################################################################
%%%%%%%%%%%%%%%%%%%%%%%%%%%%%%%%%%%%%%%%%%%%%%%%%%%%%%%%
%                    FIGURA perfiles m=0.214
%%%%%%%%%%%%%%%%%%%%%%%%%%%%%%%%%%%%%%%%%%%%%%%%%%%%%%%%
\begin{figure}[h!]
\includegraphics[width=9cm]{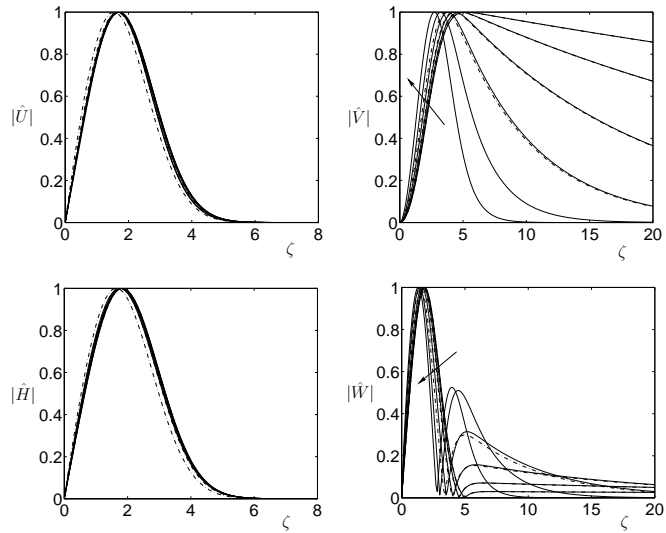}
\caption{Streamwise, cross flow velocity profiles,
and $\hat H$-profile for $\beta=0.353$
  (that value of $\beta$  considered
  in Ref.~\cite{TuminAshpis2003})
  rescaled with their maxima in $0<\zeta<\infty$, at
  $\hat x=10^{-5}, 10^{-4},10^{-3},10^{-2},10^{-1}$ and
   $1$; arrows indicate increasing values of $\hat x$.
  In addition, the asymptotic profiles, reconstructed
  using (\ref{UVHPx0}) and
 (\ref{mT1})-(\ref{mT5}), with $\lambda=\lambda_1^T$, for
 $\hat x=10^{-5}, 10^{-4},10^{-3}$, and $10^{-2}$
  are plotted with dot-dashed lines using the same rescaling.
  \label{fig3}}
\end{figure}
%%%%%%%%%%%%%%%%%%%%%%%%%%%%%%%%%%%%%%%%%%%%%%%%%%%%%%%%
%%%%%%%%%%%%%%%%%%%%%%%%%%%%%%%%%%%%%%%%%%%%%%%%%%%%%%%%
%%%%%%%%%%%%%%%%%%%%%%%%%%%%%%%%%%%%%%%%%%%%%%%%%%%

  Integration from
 $\hat x=\hat x_0=10^{-9}$ to $\hat x=1$ provides the
streamwise, cross flow velocities, and the $\hat H$-profile
(normalized with their maxima in $0<\zeta<\infty$)
illustrated in  Fig.\ref{fig3}
for   $\beta=0.353$ (that value considered in
Ref.\cite{TuminAshpis2003});  the results for other values of
$\beta$ are completely similar. Note that our analysis above captures
quite well the right behavior at small $\hat x$,
plotted with dot-dashed lines in Fig.\ref{fig3}.
This is
possible because
of our re-interpretation of Tumin asymptotic results in
terms of the new variable $\hat H$
and the use of (\ref{UVHPx0}).
Also note that, as occurred  in the case
of Blasius boundary layer \cite{HigueraV09},
the solution exhibits
an approximate selfsimilarity in the streamwise
velocity component and the variable $\hat H$,
which  (after rescaling with their maxima) are approximately
independent of $\hat x$  and approximately equal to each other.
This illustrates  the low-dimensional nature
of the MLBL equations in the  Falkner-Skan boundary layer.

                    %FIGURA
%%%%%%%%%%%%%%%%%%%%%%%%%%%%%%%%%%%%%%%%%%%%%%%%%%%%%%%%
\begin{figure}[h!]
\includegraphics[width=11cm]{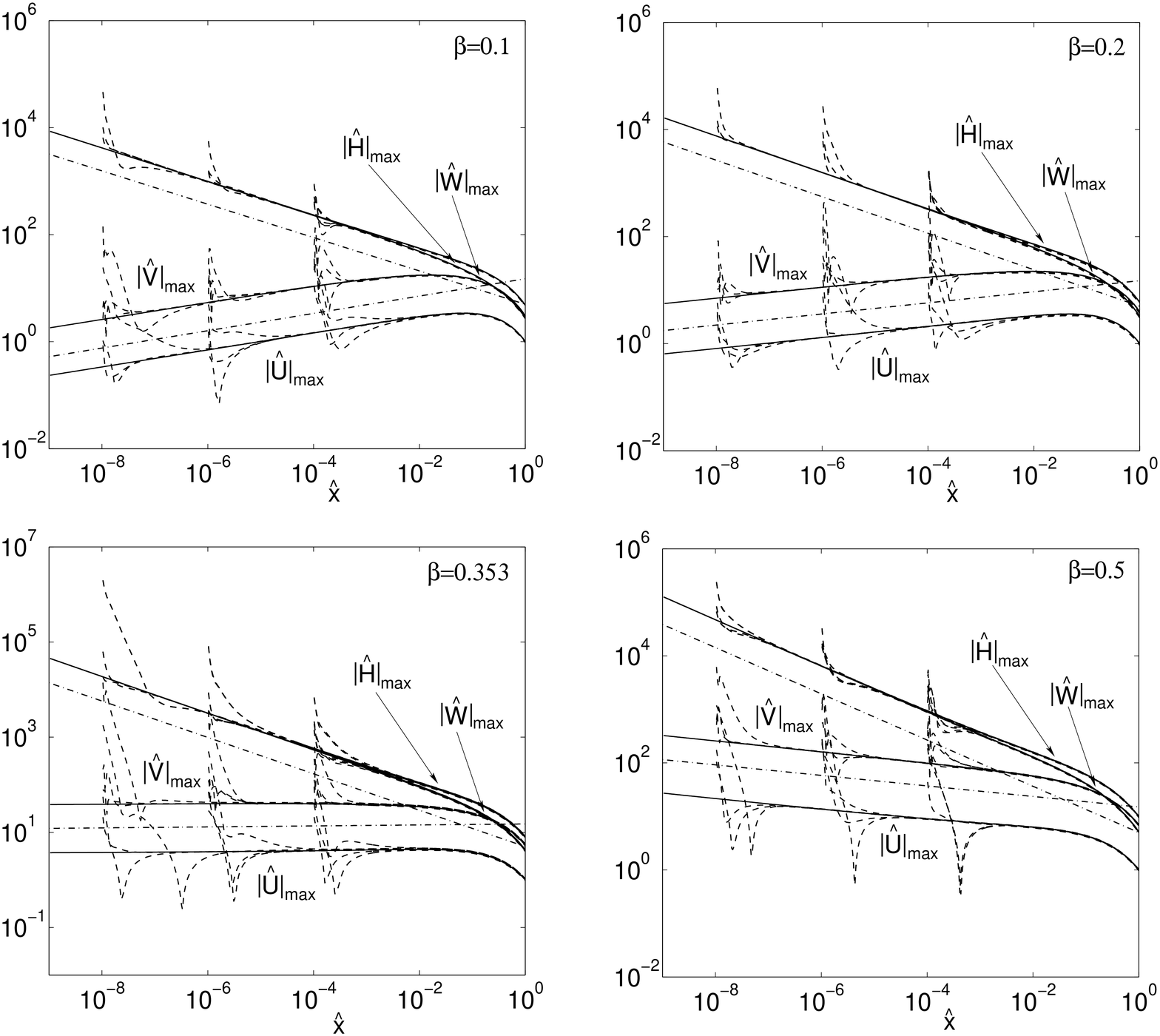}
\caption{Maxima (in $0 <\zeta <\infty$) of  $\hat U$, $\hat{V}$, $\hat{W}$,
         and $\hat H$ versus $\hat{x}$ (solid lines, as indicated by the labels),
         for $\beta=0.1,0.2,0.353$, and $0.5$. The
         asymptotic behaviors as
         $\hat x\to0$ (see section \ref{sec:leading-edge}) are
          also plotted with dot-dashed lines. The result of applying
         random initial conditions, as in (\ref{condinic}), using the
         projection on Tumin initial condition in eq.(\ref{amplmodeT}),
         is plotted with dashed lines.
\label{fig5}}
\end{figure}
%%%%%%%%%%%%%%%%%%%%%%%%%%%%%%%%%%%%%%%%%%%%%%%%%%%%%%%%
%%%%%%%%%%%%%%%%%%%%%%%%%%%%%%%%%%%%%%%%%%%%%%%%%%%%%%%%
%#################################################################
%
Figure \ref{fig5} shows with solid lines the
maxima (in $0<\zeta<\infty$) of  $\hat U$, $\hat V$,
$\hat W$, and $\hat H$ vs. $\hat{x}$ for four representative
values  $\beta$, namely $\beta=0.1,0.2, 0.353$, and 0.5.
The above mentioned scaling factor (common to all flow variables)
has been chosen such that the maximum of
 $\hat U$  vs. $\hat x$
curve in $\hat x_0<\hat x<1$ is one.
 The asymptotic behaviors near the leading
edge $\hat W \sim \hat H\sim \hat x^{\frac{1}{2-\beta}-\lambda_1^T}$
and $\hat U\sim \hat V\sim \hat x^{1-\lambda_1^T}$
(see~(\ref{UVHPx0})) are also plotted
with dot-dashed lines for comparison. Note that transient
growth is present at   $\beta=0.1,0.2$, and $0.353$,
 while everything decays
at $\beta=0.5$, as predicted in section \ref{sec:leading-edge}.

The MUSM is somewhat similar to the standard eigenmodes in
 strictly parallel flows, except of course
for the final viscous dissipation decay at large $\hat x$, which is
intrinsic to transient growth. In particular, the MUSM is the first
of  a sequence
\beqn
\textbf{q}_j=(\hat U_j, \hat V_j,\hat H_j,\hat P_j)
\text{ for }j=1,\ldots,\label{h1000}
\eeqn
which is obtained taking as initial condition at
some small $\hat x_0$ the  Tumin and Chen-Libbby eigenmodes
associated with the eigenvalues $\lambda_1,\lambda_2,\ldots$
(sorted together in increasing order). Invoking the scalings
 (\ref{UVHPx0}) and (\ref{UVHPx1}),  such initial conditions are of the form
$(0,0,\tilde H_T,0)$ and $(\tilde U_{CL},\tilde V_{CL},0,\tilde P_{CL})$
for the Tumin and Chen-Libby modes, respectively; in fact, a better
(but asymptotically equivalent) definition of the initial conditions
are obtained using  (\ref{UVHPx0}) and (\ref{UVHPx1})
with the remaining components of the Tumin
and Chen-Libby modes that have been set to zero above.
The streamwise evolving modes (\ref{h1000}) satisfy
\beqn
 \lVert \textbf{q}_1 \rVert\gg \lVert \textbf{q}_2
 \rVert\gg \dots\quad\text{in }\hat x_0\leq\hat x\leq1,\label{h1003}
 \eeqn
 provided that  $\hat x_0\ll1$. In addition,
  any streak can be expanded in series of these modes because
 any initial condition can be expanded in a series of
 the Tumin and Chen-Libby  modes.

 All these imply that after a streamwise transient any streak
 converges to the MUSM,  namely $\textbf{q}_1$.
  In addition, the projection of any streak into
  $\textbf{q}_1$ is obtained projecting its initial condition
  into the eigenfunction of the first Tumin mode,
  as given by eq.(\ref{amplmodeT}).
 This is
 illustrated in Fig.\ref{fig5}, where various
 streaks are considered that result from imposing
 at various $\hat x=\hat x_0$ random initial conditions
 of the type
$(\hat U, \hat V,\hat H,\hat P)=(\hat U_0,0,
\hat H_0, 0)$, with
\begin{equation}
\hat U_0=\hat x_0^{1-\lambda_T^1}\zeta^2F''\sum_{k=0}^4\gamma_k^1\sin k\zeta,
\quad \hat H_0=\hat x_0^{\frac{1}{2-\beta}-\lambda_T^1}\zeta^2F''\sum_{k=1}^4\gamma_k^2\cos k\zeta.
\label{condinic}
\end{equation}

Here, the coefficients  $\gamma_k^1$ and $\gamma_k^2$  are
chosen randomly between 1 and -1. Note also that fixing two
of the four variables makes sense since the initial condition should
 satisfy two compatibility conditions
 (one obtained multiplying (\ref{mlbl1}) by $F'$ and subtracting (\ref{mlbl2}),
and the second one, substituting (\ref{mlbl3})-(\ref{mlbl4}) and the
$\hat x$ derivative of (\ref{mlbl2}) into (\ref{mlbl1})),
which are selected by the equations after a
few integration steps if not satisfy initially.
These initial conditions are applied several times
for different values of  $\hat x_0$. Results (after rescaling)
are shown  with dashed lines in Fig.~\ref{fig5}.
Note that transients survive until
$\hat x/\hat x_0\sim 10$, assuming that the initial amplitudes of first
and second Tumin modes are comparable.  This is explained noting that
decaying to the MUSM is associated with the second
Tumin  mode (see Fig.\ref{fig2}), which compared with
the dominant behavior decays as
$(\hat x_0/\hat x)^\mu$, where $\mu=\lambda_1^T-\lambda_2^T=0.96, 1.02, 1.13$,
and 1.25 for
$\beta=0.1,0.2,0.353$, and $0.5$, respectively.

\section{Optimal internal streaks for $\beta<1$ \label{sec:optimalstreaks}}
%#################################################################
%#################################################################
 Optimal streaks are those streaks that show  the maximum amplification.
Finding optimal streaks is made  maximizing the  perturbed kinetic
energy gain $G \equiv E^{in} / E^{out}$ between
a generic section  $x=x_{in}$ and $x=1$,  which invoking (\ref{expansion}) is written as
\begin{equation}
 G_\text{max}=\text{max}\frac{\left[\displaystyle
 \int_0^{\infty} U^2 dy \right]_{x=1}}
          {\left[\displaystyle \text{Re}\int_0^{\infty} U^2 dy
          +\frac{g(x_\text{in})^2}{x_\text{in}^2}\int_0^{\infty} (V^2+W^2) dy \right]_{x=x_\text{in}}}.
\label{GL}
\end{equation}
A factor Re is omitted in the numerator, where a $O(1/\text{Re})$ term
depending on the cross flow velocities
is also neglected. The latter cannot be done in the denominator because,
in fact, the $O(Re)$-term becomes negligible in the maximizers.
Thus, following Luchini\cite{Luchini2000},
Tumin and Ashphis\cite{TuminAshpis2003} set to zero that term
imposing the additional condition that $\hat U=0$ at
$x_\text{in}$. Here instead, we shall retain
the $O(Re)$-term in the denominator, checking
(as in Ref.\cite{AnderssonBH99}) that results are
independent of Re provided that Re be large.

Note that maximizers of (\ref{GL}) depend on $x_\text{in}$,
with the asymptotic result as
$x_\text{in}\to0$ being most relevant.
Invoking the scaling
(\ref{b19}) and the definition (\ref{vH}), eq.(\ref{GL}) is rewritten as
\begin{equation}
 G_\text{max}(\alpha)=\text{max}
                   \frac{x_\text{in}^{\frac{\beta+1}{2-\beta}}\left[\displaystyle
                   \int_0^{\infty} \hat{U}^2 d\zeta \right]_{\hat{x}=
           \alpha^{\frac{2-\beta}{1-\beta}}}}
           {\left[\displaystyle  x_\text{in}^{\frac{2}{2-\beta}}\text{Re}\int_0^{\infty}
           \hat{U}^2 d\zeta+ \int_0^{\infty} [\hat{V}^2+(\hat H-\hat{V})^2]
           d\zeta \right]_{\hat{x}=\alpha^{\frac{2-\beta}{1-\beta}} x_\text{in}}}.
\label{GLhat}
\end{equation}

 This quotient could be maximized with the method used in
Refs.\cite{AnderssonBH99,Luchini2000,TuminAshpis2003}, who
wrote the quotient as
$(\textbf{q}_\text{out}^\top \cdot\textbf{Q}_{\text{out}} \cdot
\textbf{q}_{\text{out}})/(\textbf{q}_\text{in}^\top
\cdot\textbf{Q}_{\text{in}} \cdot\textbf{q}_{\text{in}})$,
where $\textbf{q}$ is the flow state vector
 $\textbf{q}=(U, V,H, P)$.
Using the action $\textbf{U}$ associated
with the streamwise evolution dynamical system associated with
 the MLBL equations, defined as $\textbf{q}_{\text{out}}=
\textbf{U}\cdot\textbf{q}_{\text{in}}$, and its
 upstreamwise-evolving adjoint,
 defined  such that $(\textbf{U}\cdot\textbf{q}_{\text{in}})^\top
 =\textbf{q}_{\text{in}}^\top\cdot\textbf{U}^*$,
 the quotient
  (\ref{GL}) is rewritten as
  $G=(\textbf{q}_\text{in}^\top\cdot \textbf{U}^*
  \cdot\textbf{Q}_\text{out}\cdot\textbf{U}
  \cdot\textbf{q}_\text{in})/
  (\textbf{q}_\text{in}^\top\cdot \textbf{Q}_\text{in}\cdot
    \textbf{q}_\text{in})$.
  Maximizing this leads to the
  generalized eigenvalue problem $\textbf{U}^*
  \cdot\textbf{Q}_\text{out}\cdot\textbf{U} \cdot \textbf{q}_\text{in}=G
  \textbf{Q}_\text{in}\cdot\textbf{q}_\text{in} $,
  whose maximum eigenvalue provides the maximum gain and
  the associated eigenfunctions, the maximizers.
  This problem is iteratively solved as $\textbf{q}_{\text{in},n+1}=\textbf{Q}_\text{in}^{-1}\cdot\textbf{U}^*
  \cdot\textbf{Q}_\text{out}\cdot\textbf{U} \cdot \textbf{q}_{\text{in},n} $,
   which converges quite fast provided
  that the two first eigenvalues of the above
  mentioned generalized eigenvalue problem are not too close each other.

  \subsection{A simple optimization method to calculate optimal streaks \label{sec:method}}
  The main difficulty in the method outlined above is the need of using
  the adjoint of the MLBL equations, which was made in
  Refs.\cite{AnderssonBH99,Luchini2000,TuminAshpis2003}
  for the original linearized
  boundary layer equations, written in terms of the
  unscaled wall normal coordinate $y$.
  A similar adjoint equations could be derived
  for the MLBL equations in this paper, but doing that in
  an efficient way would require to analyze the asymptotic behaviors of these
  equations as $\zeta\to\infty$ and as $\hat x\to0$. Instead,
  we propose here a simpler method that is new in this context to our knowledge.
  This method
  is based on the  observation above that the
  infinitely many solutions of the MLBL equations can be classified as a series of
  streamwise evolving `modes'  indicated
  in (\ref{h1000}) and that they satisfy (\ref{h1003}). Thus,
   if a particular solution is written as linear combination of the first $n$ such
  solutions, the resulting error scales with
  $(\hat x_\text{in}/\hat x)^{\lambda_n/\lambda_1}$, which becomes exponentially
  small as $\hat x/\hat x_\text{in}$
  is moderately large if $\lambda_n/\lambda_1$ is moderately large.
  This means that higher order modes can be safely neglected
  when optimizing (\ref{GLhat}) because these will contribute to
  increase the denominator of (\ref{GLhat}) but will have a negligible effect
  in the numerator. Now, the sequence $\lambda_1,\lambda_2,\ldots$
  increases rapidly, meaning that
  retaining a few modes would be enough to define well generic
  solutions of the MLBL equations. Thus, the method we propose to maximize
  (\ref{GLhat}) is as follows:
  \begin{enumerate}
  \item[Step 1.] Take the first $n$ Tumin  and Chen-Libby modes,
  ordered such that the associated eigenvalues are sorted in
  increasing order, as explained right after eq.(\ref{h1000}).
      \item[Step 2.] Calculate the $n$ solutions of the MLBL equations,
      \beqn
      (\hat U_1,\hat V_1,\hat H_1,\hat P_1),\ldots,
      (\hat U_n,\hat V_n,\hat H_n,\hat P_n).
      \label{h2000}
      \eeqn
      obtained taking as initial conditions at $\hat x=\hat{x}_\text{in}=
      \alpha^{\frac{2-\beta}{1-\beta}} x_\text{in}$ the $n$ modes calculated in step 1.
       \item[Step 3.] Replace the expansion
       \beqn
       (\hat U,\hat V,\hat H,\hat P)=\sum_{j=1}^na_j(\hat U_j,\hat V_j,\hat H_j,\hat P_j),
       \label{h2002}
       \eeqn
       into (\ref{GLhat}), to rewrite the quotient appearing in this equation as
       \beqn
      G_\text{max}(\alpha)=\frac{\sum_{j,k=1}^nx_\text{in}^{\frac{\beta+1}
       {2-\beta}}E^\text{out}_{jk}a_ja_k}
       {\sum_{j,k=1}^n\left(x_\text{in}^{\frac{2}{2-\beta}}
       \text{Re}E^{\text{in}1}_{jk}
       +E^{\text{in}2}_{jk}\right)a_ja_k},\label{h2004}
       \eeqn
       where  $E^\text{out}_{jk}=\int_0^{\infty} \hat U_j\hat U_k d\zeta$
        at   $\hat{x}=\alpha^{\frac{2-\beta}{1-\beta}}$, $E^{\text{in}1}_{jk}=
           \int_0^{\infty} \hat U_j\hat U_k d\zeta$ at $\hat{x}=
          \alpha^{\frac{2-\beta}{1-\beta}}x_\text{in}$, and  $E^{\text{in}2}_{jk}=
           \int_0^{\infty} [\hat V_j\hat V_k+
           (\hat H_j-\hat V_j)(\hat H_k-\hat V_k)] d\zeta$ at  $\hat{x}=
          \alpha^{\frac{2-\beta}{1-\beta}}x_\text{in}$.
           \item[Step 4.] Equation (\ref{h2004}) is the ratio of two
           $n$-th order quadratic forms, which is maximized as usually,
           solving the generalized eigenvalue problem
               \beqn
                x_\text{in}^{\frac{\beta+1}{2-\beta}}
                 \sum_{k=1}^nE^\text{out}_{jk}a_k=G
       \sum_{k=1}^n\left(x_{in}^{ \frac{2}{2-\beta}}\text{Re}E^{\text{in}1}_{jk}
       +E^{\text{in}2}_{jk}\right)a_k,\label{h2006}
               \eeqn
               whose maximum eigenvalue  $G_\text{max}$ provides
               the maximum gain in (\ref{GLhat}); the eigenvector yields
               the maximizer of (\ref{GLhat}) using (\ref{h2002}).
  \end{enumerate}

  %%%%%%%%%%%%%%%%%%%%%%%%%%%%%%%%%%%%%%%%%%%%%%%%%%%%%%%%
%                    FIGURA
%%%%%%%%%%%%%%%%%%%%%%%%%%%%%%%%%%%%%%%%%%%%%%%%%%%%%%%%
\begin{figure}[h!]
\center{
\includegraphics[width=9cm]{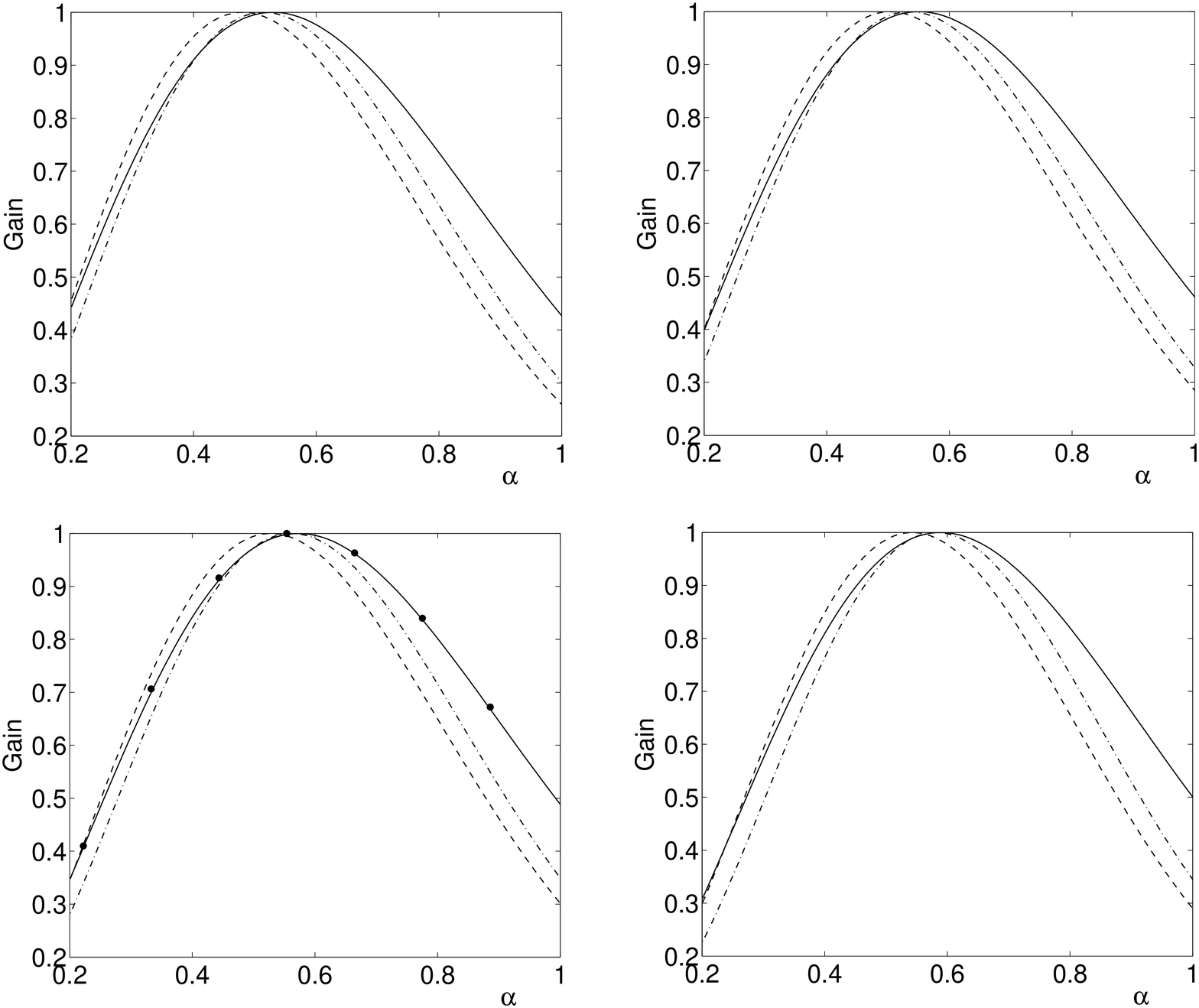}}
\caption{Rescaled (with the maximum in $\alpha$) maximun perturbed
energy gain (\ref{GLhat})  for $\beta=0.1$ (top, left), 0.2 (top, right),
$0.353$ (bottom, left), and 0.5 (bottom, right), taking the initial stage at
 $x_\text{in}=10^{-5}$  (dot-dashed lines), $10^{-2}$ (dashed lines),
 and $0.25$ (solid lines).
Results from
Ref.\cite{TuminAshpis2003} for $\beta=0.353$ and $x_\text{in}=0.25$
are plotted with plain circles.
\label{fig7}}
\end{figure}
%%%%%%%%%%%%%%%%%%%%%%%%%%%%%%%%%%%%%%%%%%%%%%%%%%%%%%%%
%%%%%%%%%%%%%%%%%%%%%%%%%%%%%%%%%%%%%%%%%%%%%%%%%%%%%%%%
   The method is tested considering the cases $\beta=0.1,0.2,0.353$
   (the case considered in Ref.\cite{TuminAshpis2003}), and $0.5$, with
   three values of the initial streamwise stage,  namely
     $x_\text{in}=10^{-5}$, $10^{-2}$, and
     0.25 (as in Ref.\cite{TuminAshpis2003}). The associated rescaled maximum
     perturbed energy gain
     curves obtained applying the method described
     above are  given in Fig.\ref{fig7}; in order to facilitate comparison for
the various values of $x_\text{in}$, the energy gain is rescaled with
its optimal value (plotted vs. $\beta$ in Fig.\ref{fig9} below).
  One, three, and five modes are enough for
  $x_\text{in}=10^{-5}, 10^{-2}$, and $0.25$, respectively,
  at the four considered values of $\beta$; retaining
  more modes provides results that
  are plot indistinguishable. As a reference, the counterpart
  obtained in Ref.\cite{TuminAshpis2003} at
  $\beta=0.353$ and $x_\text{in}=0.25$ are plotted with
  plain circles; in fact, Tumin and Ashpis took $x_\text{in}=0.111$ and
  $x_\text{out}=0.444$, which must be compared with $x_\text{in}=0.25$ and
  $x_\text{out}=1$ with the adimensionalization in the present paper.

This method provides the maximum gain in a quite fast and robust way.
Note that the method does not require any calculation of adjoint
equations. Instead, only a few  Tumin and Chen-Libby eigenmodes
are needed. But these are not really necessary, noting that the
$n$ solutions of the MLBL equations considered in step 2 can be
replaced in the method by any set of $n$ solutions with initial
conditions at $\hat x=\hat x_\text{in}\equiv
           \alpha^{\frac{2-\beta}{1-\beta}}x_\text{in}$ that are linearly
           independent and exhibit a significant projection
           into the $n$ first Tumin and Chen-Libby eigenmodes. For instance, we can
integrate the MLBL equations taking at some smaller value of $\hat x$
(say, $\hat x_0=\hat x_\text{in}/10$) the following $n$
initial conditions (cf (\ref{condinic}))
\begin{equation}
(\hat U_k,\hat V_k, \hat H_k,\hat P_k)=\left(\hat x_0^{1-\lambda_T^1}\zeta^2F''\sin k\zeta,0,
\hat x_0^{\frac{1}{2-\beta}-\lambda_T^1}\zeta^2F''\cos k\zeta,0\right),
\label{h2010}
\end{equation}
for $k=1,\ldots,n$, and considering in step 2 these solutions
 for $\hat x\geq\hat x_\text{in}$. The resulting modified  method,
 retaining the same numbers of  modes as above
 produce the same  results
 in  Fig.\ref{fig7} to plot accuracy. Since this
 modification provides a simpler method, it is the
 resulting modified method that we propose in this paper.

  Note that  the number of required modes in Fig.\ref{fig7} decreases as
 $x_\text{in}$ decreases. In particular, just one mode is enough if
 $x_\text{in}\leq10^{-5}$, and furthermore, the resulting maximizer
 coincides with the MUSM defined in section \ref{sec:usm}, as was to be expected.
 In fact, taking into account the behavior of the MUSM as $\hat x\to0$
 (given by the eigenfunction associated with the first Tumin eigenvalue),
 the following asymptotic value of the gain (\ref{GLhat}) is readily
 obtained that does not require any maximization process
 \begin{equation}
  G_\text{max}(\alpha)=\frac{x_{in}^{\frac{\beta+1}{2-\beta}}
 \left[\displaystyle
                   \int_0^{\infty} \hat{U}^2 d\zeta \right]_{\hat{x}=
           \alpha^{\frac{2-\beta}{1-\beta}}}}
           {\left[\displaystyle   \int_0^{\infty} \hat H^2
           d\zeta \right]_{\hat{x}=\alpha^{\frac{2-\beta}{1-\beta}} x_{in}}}\sim
           x_\text{in}^{2\lambda_T^1+\frac{\beta-1}{2-\beta}}\quad\text{as }x_\text{in}\to0.
\label{h3000}
\end{equation}
Here, $\hat U$ and $\hat H$ are the associated components of
the MUSM, at $\hat{x}=\alpha^{\frac{2-\beta}{1-\beta}}$
and $\hat{x}=\alpha^{\frac{2-\beta}{1-\beta}} x_\text{in}$,
respectively. The indicated asymptotic behavior as $x_\text{in}\to0$ results
from the scaling (\ref{UVHPx0}) of the first Tumin mode,
associated with the eigenvalue $\lambda_T^1$.

%%%%%%%%%%%%%%%%%%%%%%%%%%%%%%%%%%%%%%%%%%%%%%%%%%%%%%%%
%                    FIGURA
%%%%%%%%%%%%%%%%%%%%%%%%%%%%%%%%%%%%%%%%%%%%%%%%%%%%%%%%
\begin{figure}[h!]
\includegraphics[width=6cm,height=4cm]%{Curva_ALPHAopt_vs_Beta_x01e-9_x01e-18.eps}
{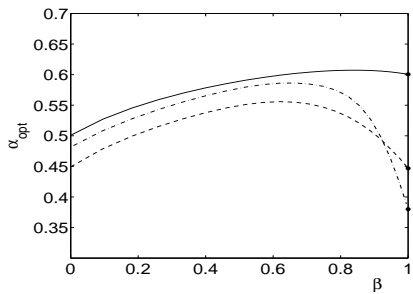}
\caption{Optimal  spanwise wavenumber $\alpha_\text{opt}$
 vs. $\beta$ at $x_\text{in}=10^{-5}$ (dot-dashed line), $10^{-2}$ (dashed line), and 0.25
 (solid line). Plain circles at
$\beta=1$ result from calculations in the Appendix.
\label{fig9}}
\end{figure}
%%%%%%%%%%%%%%%%%%%%%%%%%%%%%%%%%%%%%%%%%%%%%%%%%%%%%%%%%
%%%%%%%%%%%%%%%%%%%%%%%%%%%%%%%%%%%%%%%%%%%%%%%%%%%%%%%%%
To complete the results above, the optimal gain and the
associated value of the spanwise wavenumber
 are plotted vs. $\beta$ in Fig.\ref{fig9} for
$x_\text{in}=10^{-3},10^{-2}$, and 0.25. The limiting values at $\beta=1$ result from
the analysis in the Appendix, at the end of the paper. Note that $\alpha_\text{opt}$  depends
non-monotonously both on $\beta$ and on $x_\text{in}$.

Summarizing,   a simple optimization method has been proposed to
calculate optimal streaks that does  not rely on the
adjoint formulation. Instead, minimization
is made on the low dimensional manifold spanned by a
few solutions of the direct problem. In fact, five solutions
are enough in the streaks of the Falkner-Skan boundary layer for
$x_\text{in}\leq0.25$.
In addition, only one such solution is enough for small
$x_\text{in}$ ($x_\text{in}\leq10^{-5}$), meaning that no
optimization process is necessary in this limit. We believe
that the method proposed above is also useful  to treat related
transient growth problems in Fluid Dynamics, since the main
 ingredient that allowed constructing such method is usual
in these problems. This ingredient is that the behavior in
the growth stage of transient growth are described by an
eigenvalue problem (Tumin problem in the case considered
in this paper), whose eigenvalues are somewhat  separated.
And furthermore, the method could work in nonlinear,
time dependent parabolic problems as well, since the
dynamics of parabolic problems is low dimensional at large time\cite{FoiasST88}.

%#################################################################
%#################################################################
\section{Concluding remarks \label{sec:Concluding}}
%#################################################################
%#################################################################
 An analysis of optimal streaks in the Falkner-Skan
 boundary layer has been performed
concentrating in various issues that are now
summarized, and are expected to apply to related transient growth
boundary layer problems:
\begin{itemize}
\item The careful analysis of the behavior of
streaks near the free stream was necessary to
 construct a quite efficient numerical
scheme, which allowed  for integrating from extremely small values
of the streamwise coordinate. The analysis of the free stream behavior also
allowed for making the
correct interpretation of the behavior near the leading
edge, in terms of a new variable $\hat H$ that behaves
as the streamwise vorticity near the leading edge.
\item The already known approximate selfsimilarity
of the solution has been completed in terms of the
new variable $\hat H$.
\item Streaks behave as `modes', which can be
classified according to their behavior near the leading edge.
 The analogy with standard modes in strictly parallel flows
becomes clear comparing the cases $\beta<1$ (considered in most
part of the paper) and $\beta=1$ (considered in the Appendix).
The most dangerous mode was called the
MUSM and played an essential role in understanding
the streamwise evolution of streaks.
\item A  simple optimization method has been proposed
that does not rely on the adjoint equations and allows
for the fast and precise computation of optimal streaks.
    \item Optimal streaks can be directly defined
    from the MUSM, without the need of any
    optimization process, if the initial stage is sufficiently
     close to the leading edge.
        \item All the above is somehow a pre-requisite
         to derive simple, yet sufficiently precise descriptions
        of the interaction between longitudinal streaks
        and transversal Tollmien-Schlichting modes,
        which can be still considered a major open problem in the field.
        \item The analysis above applies to internal streaks,
        resulting from, e.g., obstacles near the leading edge.
         External streaks forced by perturbations in the outer
         stream show a different behavior near the leading
         edge and require a different treatment, which is
         currently under research.
\end{itemize}

%\bibliography{boundary_layer}

\vskip1cm

\appendix
{\bf APPENDIX: THE CASE $\beta\geq1$}\\

In the case $\beta=1$ the spanwise wavenumber $\alpha$ cannot be eliminated from
the LBL equations (\ref{mlbl10})- (\ref{mlbl40}), but the
right hand sides of these are independent of $x$ and thus
can be expanded in terms of normal modes as (cf eq.(\ref{UVHPx0}))
\beqn
(U,V,W,P)=\sum_{j=1}^\infty a_j(\tilde U_j,\tilde V_j,
\tilde W_j,\tilde P_j)\hat x^{1-\lambda_j}+\cc,\label{ap1}
\eeqn
where {\cc}  stands for the complex conjugate and the modes and exponents
are given by the following eigenvalue problem
\begin{alignat}{1}
&(1-\lambda)\tilde{U}+ \tilde V'-\alpha\tilde W=0,\label{ap3} \\
&\tilde U'' +  F\tilde U'+
[(\lambda-2) F'-\alpha^2]\tilde U
-F''\tilde V=0,\label{ap5}\\
& \tilde V''+F\tilde V'+
[(\lambda+1)F'-\alpha^2]V=\tilde P',\label{ap7}\\
&\tilde W''+F\tilde W'+
(\lambda F'-\alpha^2)\tilde W
=\alpha \tilde P,\label{ap9}\\
&\tilde U=\tilde V= \tilde W =0 \quad\text {at}\quad \zeta=0\text{ and }
\infty,\quad  \tilde P=0\quad\text {at}\quad  \zeta=\infty.\label{ap11}
\end{alignat}
Note that as $\alpha\to0$ two scalings are possible. Rescaling $\tilde W$ as
$\tilde W/\alpha$ and setting $\alpha=0$, (\ref{ap3}) and (\ref{ap9}) lead to
the Tumin eigenvalue problem (\ref{mT1})-(\ref{mT4}), while setting $\alpha=0$ in
(\ref{ap3}) and (\ref{ap11}), the Chen-Libby eigenvalue problem (\ref{mT3})-(\ref{mT3a}),
(\ref{ChL1}), and (\ref{ChL4}) results. This  means that the limiting values of the eigenvalues
of (\ref{ap3}) and (\ref{ap11}) as $\alpha\to0$ are Tumin and Chen-Libby eigenvalues.

\begin{figure}[h!]
\includegraphics[width=5cm]{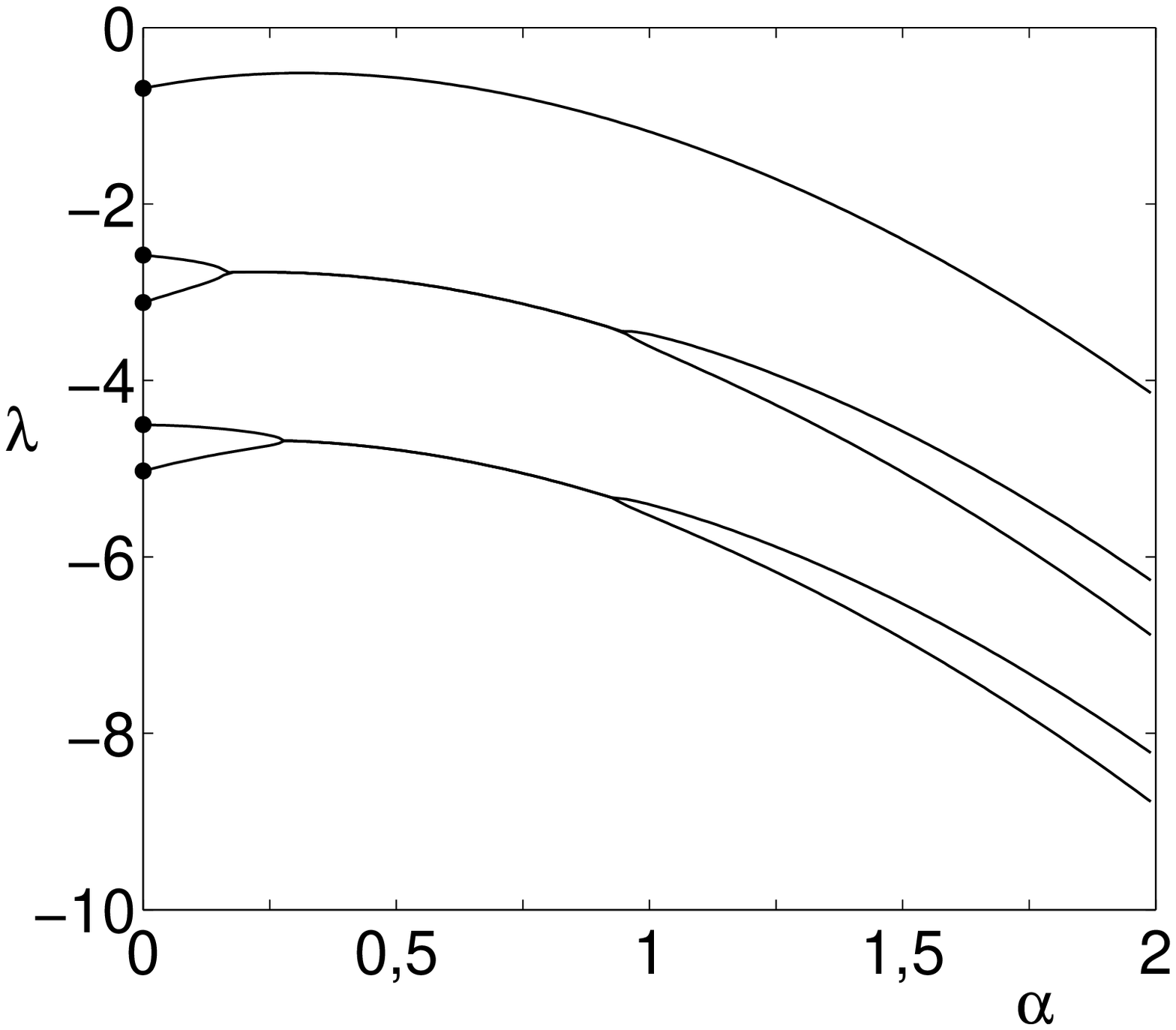}\includegraphics[width=5cm]{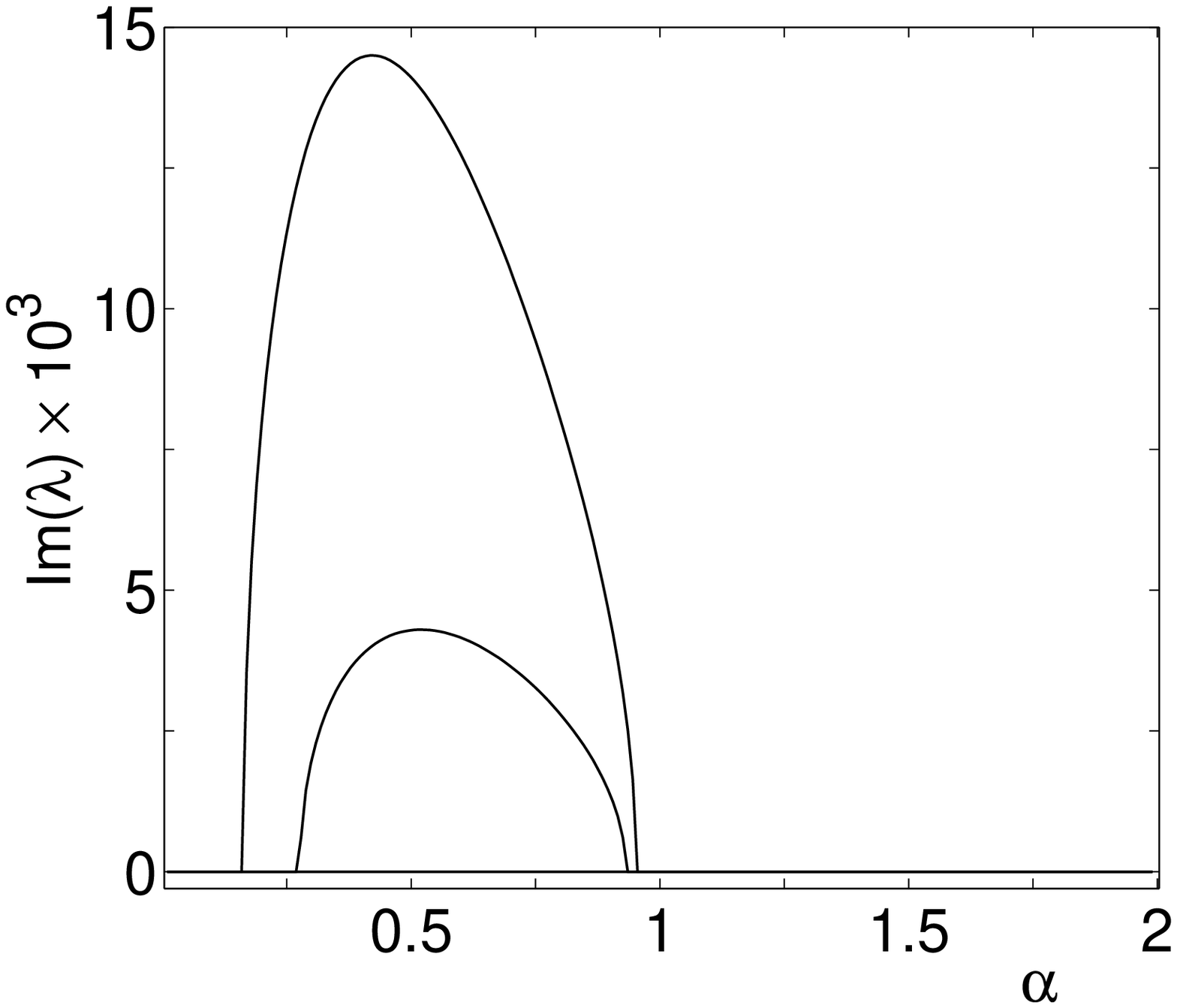}
\caption{The real (left) and imaginary (right) parts of the first
three eigenvalues  of (\ref{ap3})-(\ref{ap11}) vs. $\alpha$. Plain circles at
$\alpha=0$ correspond to the Tumin and Chen-Libby eigenvalues at $\beta=1$, plotted
with plain circles in Fig.\ref{fig2}. \label{fig10}}
\end{figure}
The eigenvalues of (\ref{ap3}) and (\ref{ap11}) are generally complex, as shown in Fig.\ref{fig10},
where the real and imaginary parts of first
three eigenvalues are plotted vs. $\alpha$; note that the second
and third eigenvalues are complex in the intervals $0.16<\alpha<0.96$ and
$0.27<\alpha<0.94$,  respectively.

The limiting values as $\beta\to1$ of the optimal wavenumbers in Fig.\ref{fig9} are
 calculated minimizing the counterpart of the ratio
(\ref{h2004}), using $n$ of the modes (\ref{ap1}), which is
 \beqn
       G_\text{max}(\alpha)=\frac{\sum_{j,k=1}^nx_{in}^2E^\text{out}_{jk}\bar a_ja_k}
       {\sum_{j,k=1}^n\left(x_{in}^2
       \text{Re}E^{\text{in}1}_{jk}
       +E^{\text{in}2}_{jk}\right)\bar a_j a_k},\label{ap13}
       \eeqn
       where overbar stands for the complex conjugate,
         $E^\text{out}_{jk}=\int_0^{\infty} \bar{\tilde U}_j\tilde U_k d\zeta$
        at   $\hat{x}=\alpha$, $E^{\text{in}1}_{jk}=
           \int_0^{\infty} \bar{\tilde U}_j\tilde U_k d\zeta$ at
           $\hat{x}=
          \alpha x_\text{in}$, and  $E^{\text{in}2}_{jk}=
           \int_0^{\infty} [\bar{\tilde V}_j\tilde V_k+
           (\bar{\tilde H}_j-\bar{\tilde V}_j)(\tilde H_k-\tilde V_k)] d\zeta$
           at  $\hat{x}=
          \alpha x_\text{in}$. The resulting optimal
          gain calculated from the generalized eigenvalue problem (cf (\ref{h2006}))
          \beqn
          \sum_{k=1}^nx_{in}^2E^\text{out}_{jk}a_k=
          G\sum_{k=1}^n\left(x_{in}^2
       \text{Re}E^{\text{in}1}_{jk}
       +E^{\text{in}2}_{jk}\right) a_k,\label{ap15}
       \eeqn
       whose maximum eigenvalue $G_\text{max}$ provides the maximum of the energy gain
       (\ref{ap15}). Note that although the amplitudes $a_k$ are complex, the eigenvalues
       are real. This is because the matrices appearing in (\ref{ap15}) are Hermitian.
          Using these, the maximum perturbed energy gain is calculated from the
           initial stages $x_\text{in}=10^{-5}$, $10^{-2}$, and 0.25, which  peak at $\alpha=0.38$, 0.45, and 0.60,
          respectively. These  are precisely the limiting values
          of their counterparts calculated for $\beta<1$ in subsection
          \ref{sec:method} and plotted in Fig.\ref{fig9}.

\begin{figure}
\includegraphics[width=6cm]{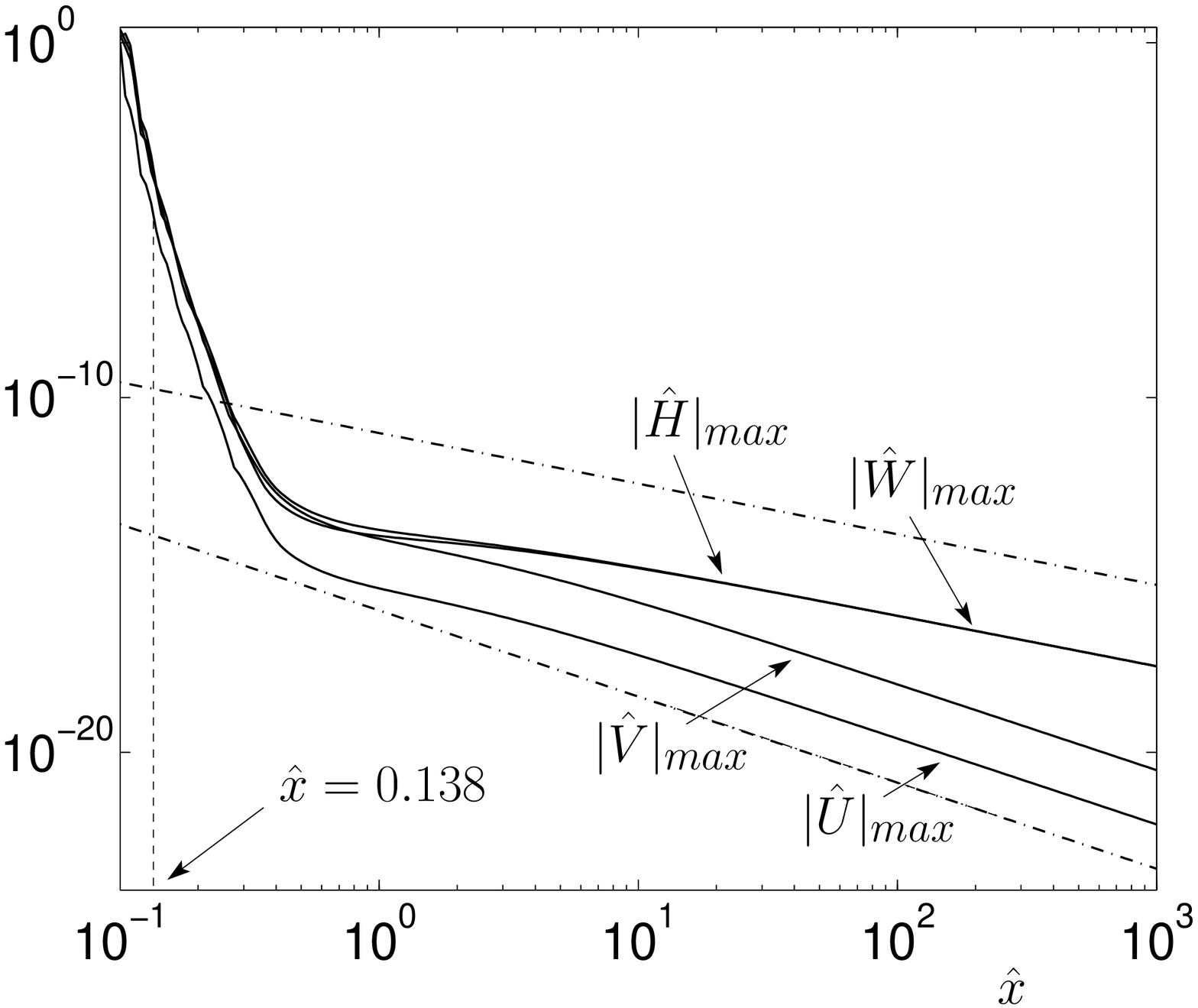}
\includegraphics[width=6cm]{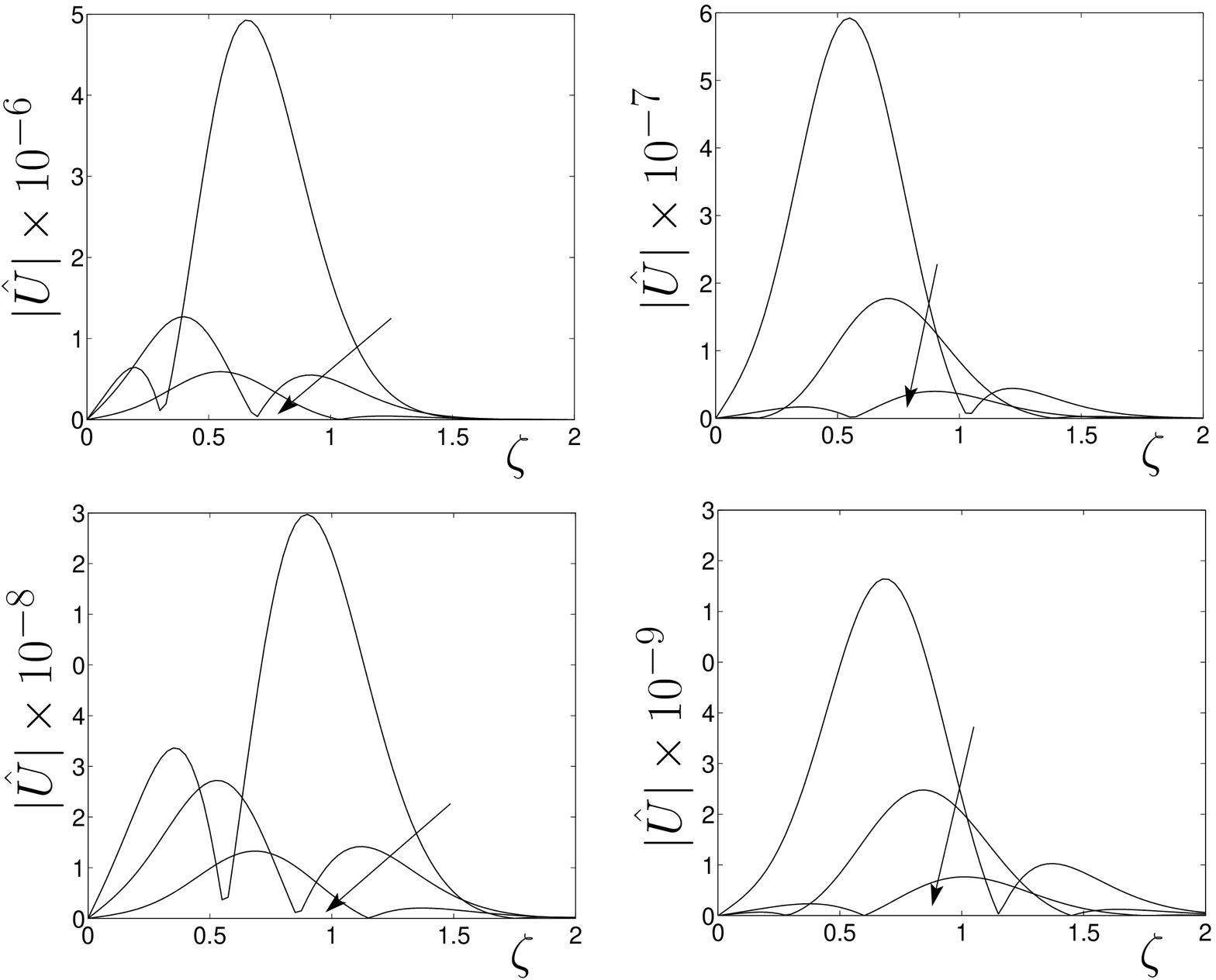}
\caption{The case $\beta=1.5$. Left: Maxima (in $0<\zeta<\infty$
 of $|\hat U|$, $|\hat V|$,
          $|\hat H|$, and $|\hat W|$ vs. $\hat x$.
          The upper and lower straight dot-dashed lines provide the asymptotic behaviors of $|H|$ and $|U|$, which invoking (\ref{b19}), (\ref{b50}), and
          (\ref{UVHPx0}) are $x^{1-\lambda}$
          and $x^{1-\lambda}/g(x)$, respectively.
          Right:
          A sequence of profiles of
          $|\hat U|$ at nine  equispaced values of $x$ between $x=0.1389$ and $0.1995$, $x_1,\ldots,x_9$.
          Due to the exponential decay, these profiles cannot be plotted in the same plot. Instead, two of them are given in each plot, which also contains the last profile in the former subplot, namely the four subplots provide the profiles at $x=x_1,x_2,x_3$ (top, left), $x=x_3,x_4,x_5$ (top, right), $x=x_5,x_6,x_7$ (bottom, left), and $x=x_7,x_8,x_9$ (top, right). Arrows
          indicate increasing values of $\hat x$.  \label{fig11}}
\end{figure}

          Even though the case $\beta>1$ does not produce transient
          growth, a brief description is given here to complete the analysis.
          This case yields a different behavior of the solutions
          of the MLBL equations (\ref{mlbl1})-(\ref{mlblcc}) as $\hat x\to0$. This
          is because $g(\hat x)=\hat x^{\frac{1-\beta}{2-\beta}}\to\infty$ as
          $\hat x\to0$ in this case. Since $g(\hat x)$ is large,
          $\hat U$, $\hat V$, $\hat H$, and $\hat P$ converge to zero exponentially fast
          as $g(\hat x)\zeta$ is large, which means that activity is concentrated
          in a thin wall-normal layer near the wall. The $\hat x$-behavior in this stage
          is given by an eigenvalue problem that is omitted here for the sake of brevity.
          As $\hat x\to\infty$ instead $g(\hat x)$ is small and the Tumin and Chen-Libby
          eigenvalue problems are recovered. Thus, the solution converges to the first
          Tumin mode as $\hat x\to\infty$. In order to illustrate all these, the
          MLBL equations are integrated for $\beta=1.5$, between
          $\hat x=\hat x_0=0.1$ and $10^3$. The maxima of $|\hat U|$, $|\hat V|$,
          $|\hat H|$, and $|\hat W|$ are plotted with solid lines in
          Fig.\ref{fig11} left. Note that all these decay quite fast in an initial
          stage, followed by a transition to the behavior given by the first
          Tumin eigenmode, plotted with dot-dashed lines. The irregular behavior at
          small $\hat x$ is not a numerical artifact, but is due to the fact that
          we are plotting maximum values, and the maxima alternated between two positions,
          as illustrated in the right plot, where the rescaled profiles of $|\hat U|$ are plotted
          for various values of $\hat x$, as indicated.

\end{document}